\documentclass[twocolumn]{aastex631}
\usepackage{mathtools}
\usepackage{dsfont}
\usepackage[nolist]{acronym} 
\usepackage[nameinlink,capitalise,noabbrev]{cleveref}

\newcommand\pc{\;{\rm pc}}
\newcommand{\REV}[1]{{{#1}}}

\makeatletter
\usepackage{etoolbox}
\patchcmd\H@refstepcounter{\protected@edef}{\protected@xdef}{}{}
\makeatother

\shorttitle{Turbulent Equilibrium Spheres}
\shortauthors{Moon \& Ostriker}

\begin{document}

\title{Theory of Turbulent Equilibrium Spheres with Power-Law Linewidth-Size Relation}

\author[0000-0002-6302-0485]{Sanghyuk Moon}
\affiliation{Department of Astrophysical Sciences, Princeton University,
  Princeton, NJ 08544, USA}
\author[0000-0002-0509-9113]{Eve C.\ Ostriker}
\affiliation{Department of Astrophysical Sciences, Princeton University,
  Princeton, NJ 08544, USA}
  \affiliation{Institute for Advanced Study, 1 Einstein Drive, Princeton, NJ 08540, USA}
\email{sanghyuk.moon@princeton.edu, eco@astro.princeton.edu}

\begin{acronym}
    \acro{BE}{Bonnor-Ebert}
    \acro{TES}{turbulent equilibrium sphere}
    \acro{SPS}{singular polytropic sphere}
    \acro{ISM}{interstellar medium}
    \acro{GMC}{giant molecular cloud}
\end{acronym}

\begin{abstract}
  Dense cores inherit turbulent motions from the interstellar medium in which they form.
  As a tool for comparison to both simulations and observations, it is valuable to construct theoretical core models that can relate their internal density and velocity structure while predicting their stability to gravitational collapse.
  To this end, we solve the angle-averaged equations of hydrodynamics under two assumptions: 1) the system is in a quasi-steady equilibrium; 2) the velocity field consists of radial bulk motion plus isotropic turbulence, with turbulent dispersion increasing as a power-law in the radius.
  The resulting turbulent equilibrium sphere (TES) solutions form a two-parameter family, characterized by the sonic radius $r_s$ and the power-law index $p$.
  The TES is equivalent to the Bonnor-Ebert (BE) sphere when $r_s\to \infty$.
  The density profile \REV{in outer regions} of the TES is slightly shallower than the BE sphere, but is steeper than the logotropic model.
  Stability analysis shows that the TESs with size exceeding a certain critical radius are unstable to radial perturbations. 
  The center-to-edge density contrast, mass, and radius of the marginally stable TES all increase with increasing average velocity dispersion.
  The FWHM of the column density profile is always smaller than the critical radius, by a larger factor at higher velocity dispersion, suggesting that observations need to probe beyond the FWHM to capture the full extent of turbulent cores.
  When applied to the highly turbulent regime typical of cluster-forming clumps, the critical mass and radius of the TES intriguingly resembles the typical mass and radius of observed star clusters.
\end{abstract}

\section{Introduction}\label{sec:intro}

Dense prestellar cores are roughly spherical, compact ($\lesssim 0.1\,\mathrm{pc}$), centrally-concentrated objects found at the bottom of the hierarchy of \ac{ISM} structure.
Due to gravity, cores are internally stratified, with 
radial profiles of density $\rho$ characterized by a central plateau surrounded by an outer envelope approximately following $\rho\propto r^{-2}$ \citep[and references therein]{bergin07}.
Some cores exhibit molecular line profiles that cannot be explained by thermal motions alone, but could be interpreted as signatures of infall or turbulence \citep[e.g.,][]{goodman98,cwlee01,tafalla04}.
While the true dynamical nature of dense cores -- individually and as a class -- is still under debate, it is clear that a fraction of them undergo gravitational collapse, leading to star formation.

The radial column density profile of some observed cores closely matches the theoretical prediction of the isothermal equilibrium known as the \ac{BE} sphere \citep{bonnor56,ebert55}, suggesting that those particular cores obey hydrostatic equilibrium, in which self-gravity is balanced entirely by a radial thermal pressure gradient \citep[e.g.,][]{alves01,tafalla04}.
Stability analyses have shown that such equilibria are unstable unless they are truncated at a small enough radius \citep{bonnor56,ebert57,lynden-bell68,sormani13}.

The mass and radius of the marginally stable \ac{BE} sphere are called the \ac{BE} mass, $M_\mathrm{BE}$, and the \ac{BE} radius, $R_\mathrm{BE}$, respectively, given by 
\begin{equation}\label{eq:mbe}
\begin{split}
  M_\mathrm{BE} &= 4.43 M_G(\rho_c),\\
  &= 1.18 M_G(\rho_e),\\
  &= 1.86 M_G(\overline{\rho}),
\end{split}    
\end{equation}
\begin{equation}\label{eq:rbe}
\begin{split}
  R_\mathrm{BE} &= 1.82 R_G(\rho_c),\\
  &= 0.486 R_G(\rho_e),\\
  &= 0.762 R_G(\overline{\rho}),
\end{split}
\end{equation}
where $\rho_c$, $\rho_e$, and $\overline{\rho}\equiv 3M_\mathrm{BE}/(4\pi R_\mathrm{BE}^3)$ are the center, edge, and average density of a critical \ac{BE} sphere, respectively, with $\rho_c/\rho_e=14.0$ and $\bar\rho/\rho_e=2.45$.
Here, $M_G$ and $R_G$ are the characteristic gravitational mass and radius for an isothermal system obtained from dimensional analysis as a function of density $\rho$, 
\begingroup
\allowdisplaybreaks
\begin{align}
  \begin{split}\label{eq:mgrav}
    M_G &\equiv \frac{c_s^3}{G^{3/2}\rho^{1/2}}\\
        &= 1.27\,M_\odot \left( \frac{T}{10\,\mathrm{K}} \right)^{3/2} \left( \frac{n_\mathrm{H}}{10^4\,\mathrm{cm}^{-3}} \right)^{-1/2},
  \end{split}\\
  \begin{split}\label{eq:rgrav}
    R_G &\equiv \frac{c_s}{G^{1/2}\rho^{1/2}}\\
        &= 0.15\,\mathrm{pc} \left( \frac{T}{10\,\mathrm{K}} \right)^{1/2} \left( \frac{n_\mathrm{H}}{10^4\,\mathrm{cm}^{-3}} \right)^{-1/2},
  \end{split}
\end{align}
\endgroup
where $c_s=(kT/\mu)^{1/2}$ is the sound speed, $G$ is the gravitational constant, and $n_\mathrm{H} = \rho/\mu_\mathrm{H}$ is the hydrogen number density, with $\mu = 2.3m_\mathrm{H}$ and $\mu_\mathrm{H} =1.4m_\mathrm{H}$ assuming $10\%$ helium abundance by number.

Although the \ac{BE} sphere provides essential physical insight regarding the onset of the core collapse, its applicability is limited because it neglects non-thermal pressure arising from internal turbulent motions, which are known to be present in real dense cores.
In particular, the observed turbulent velocity dispersion within dense cores is known to increase with distance from the core center approximately as a power-law \citep{fuller92,caselli95,choudhury21}\footnote{Although some studies indicate that the turbulent velocity dispersion flattens out near the core center, forming a ``coherent'' region \citep{goodman98,chen19}, this could be due to projection effects (see \cref{sec:projection} for related discussion).}.
In order to take into account the effect of turbulence, \citet[see also \citealt{lizano89}]{mclaughlin96,mclaughlin97} constructed a theoretical model of isothermal spheres supported by non-thermal pressure using a phenomenological, ``logotropic'' equation of state $P/P_c = 1 + A \log (\rho/\rho_c)$, where $P_c$ and $\rho_c$ are the central pressure and density, respectively, and $A$ is a free dimensionless parameter.
However, such an equation of state is not derived from first principles, and the linewidth-size relation that it predicts is inconsistent with observations except for a limited range of radius, as pointed out by \citet{mckee03}.
Other models simply replaced the pressure term in the hydrostatic equation with assumed nonthermal pressure \citep{myers92} or employed a composite polytrope \citep{curry00}.

For application to cores within molecular clouds, the
\ac{BE} sphere is also problematic in that it assumes a definite outer boundary at which the density profile is sharply truncated, which is in stark contrast to real cores that continuously merge into the ambient surrounding cloud
\footnote{Cores embedded in much warmer gas can in principle be truncated by the thermal pressure of hotter phase gas, but this is not a realistic star formation environment.}.
For example, it has been pointed out that the critical center-to-edge density contrast of $14$ is too small compared to the large dynamic range of volume density in \acp{GMC}, such that it would be extremely difficult for a core to remain stable \citep{shu87,vazquez-semadeni05}.
However, this objection does not account for the fact that a core never exists in isolation: the gravitational field away from the center of the core is increasingly governed by surrounding structures (dense filaments, other cores, and stars).
Thus, it is entirely possible that there exists an effective outer ``tidal boundary'' determined by the structure of gravitational potential around a core.

\begin{figure*}[htpb]
  \plotone{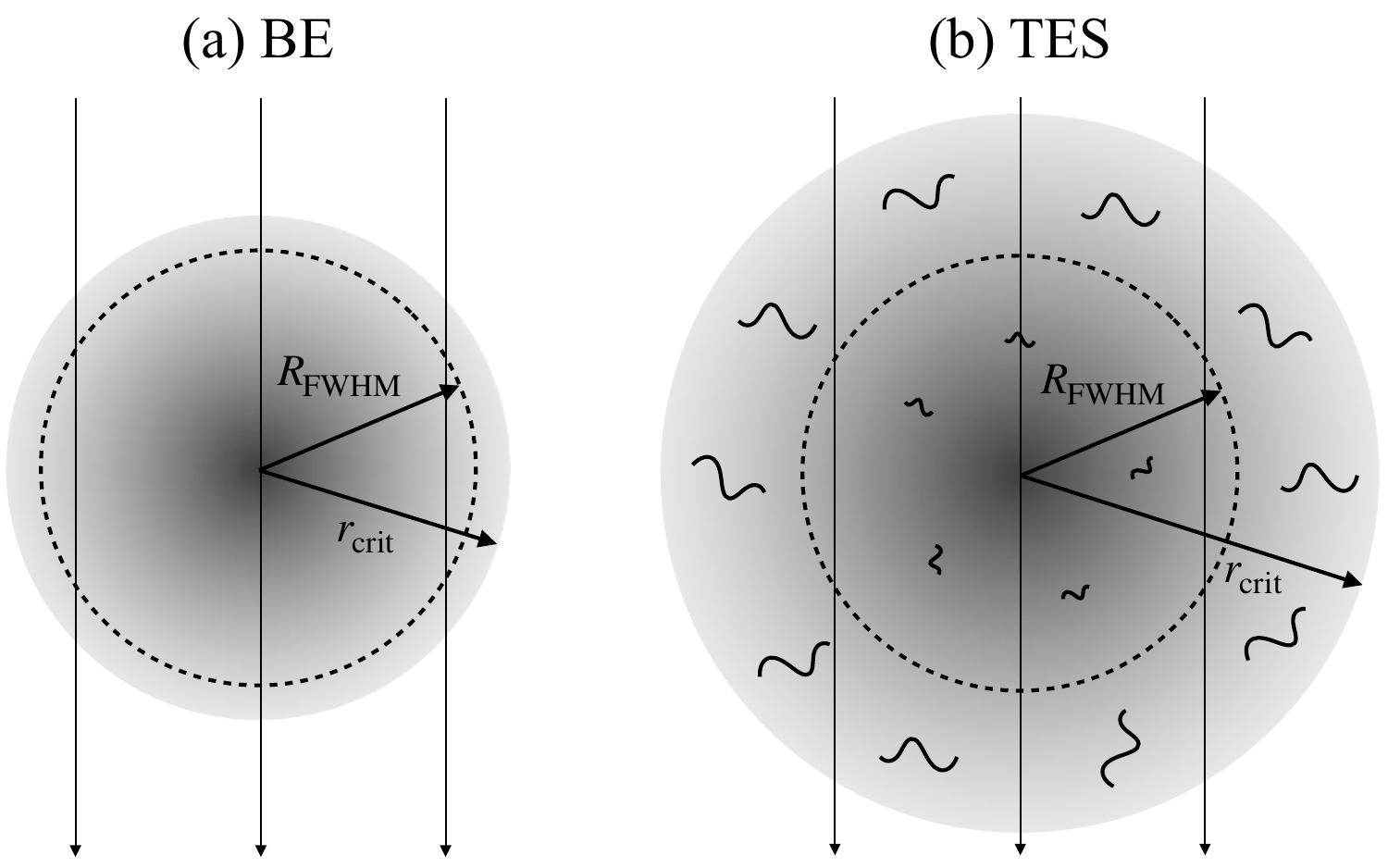}
  \caption{Schematic of (a) a \acf{BE} sphere and (b) a \acf{TES}.
  The latter is a member of a family of exact equilibrium solutions to the angle-averaged equations of hydrodynamics, in which the velocity field is decomposed into bulk and turbulent motions, where the latter is assumed to be isotropic and obey a power-law linewidth-size relationship (see \cref{sec:angle-averaged-eom,sec:equilibrium_sequence}).
  Both the \ac{BE} sphere and \ac{TES} solution families possess a critical radius $r_\mathrm{crit}$, such that for $r> r_\mathrm{crit}$ the equilibrium is unstable to radial perturbations (\cref{sec:stability}).
  The widely used observational core radius $R_\mathrm{FWHM}$ is comparable to $r_\mathrm{crit}$ only when the underlying turbulence is very weak (i.e., close to the \ac{BE} sphere), as will be shown in \cref{fig:critical_tes_obs_properties}.
  Vertical arrows represent line-of-sight projection, which shows 
  that the observed linewidth may contain a non-thermal component even at the center of the core (discussed in \cref{sec:projection}).}
  \label{fig:be_vs_tes}
\end{figure*}

The primary goal of this paper is to construct a physical model of a dense core supported by both thermal and turbulent pressure.
Instead of assuming a relationship between pressure and density, we directly solve the angle-averaged equations of hydrodynamics assuming a statistical quasi-equilibrium and a power-law linewidth-size relationship.

\cref{fig:be_vs_tes} visually illustrates the basic concept of our \acf{TES} model, juxtaposed with the traditional \ac{BE} sphere with no turbulent motion.
The results presented here quantify how the critical density contrast, radius, and mass change with the properties of underlying turbulent velocity field.
The \ac{BE} sphere naturally appears as a limiting solution with vanishing turbulent velocities.

In a companion paper (Paper II), we will show that quasi-equilibrium structures resembling the \ac{TES} emerge in three-dimensional simulations of self-gravitating, isothermal turbulence with conditions similar to star-forming GMCs.
\REV{The simulations presented in Paper II demonstrate that the size of the local potential well associated with a core is limited by the gravity from neighboring structures, imposing an effective maximum core radius.
This allows stable equilibria to exist in an isothermal medium with continuous density distribution.}
Paper II will analyze evolution of individual cores and gravitational potential structure around them, to identify critical conditions that determine the onset of the collapse.

The remainder of this paper is organized as follows.
In \cref{sec:angle-averaged-eom}, we derive the equation of motion governing the radial dynamics of a turbulent core.
In \cref{sec:equilibrium_sequence}, we introduce dimensionless variables and derive equilibrium solutions.
In \cref{sec:stability}, we perform a stability analysis to obtain the critical radius beyond which the equilibrium becomes unstable. Equilibrium solutions truncated at this radius are termed ``critical cores.''
\cref{sec:physical_conditions} presents the physical properties of critical cores as a function of the average velocity dispersion.
Finally, we summarize our work and discuss its implications in \cref{sec:discussion}.

\section{Angle-averaged Equation of Motion}\label{sec:angle-averaged-eom}

In this section, we derive the angle-averaged Lagrangian equation of motion (\cref{eq:lagrangian_eom}) satisfied by a spherical, isothermal region pervaded by turbulence.
We start by writing the continuity equation and the radial component of the momentum equation in  spherical coordinates as follows:
\begin{equation}\label{eq:continuity-spherical-polar-expansion}
\begin{split}
  \frac{\partial\rho}{\partial t}
  + \frac{1}{r^2}\frac{\partial}{\partial r} \left( r^2\rho v_r \right)
  + \frac{1}{r\sin\theta}\frac{\partial}{\partial \theta} \left( \sin\theta \rho v_\theta \right)\\
  + \frac{1}{r\sin\theta}\frac{\partial}{\partial\phi} \left( \rho v_\phi \right)
  = 0,    
\end{split}
\end{equation}
\begin{equation}\label{eq:momentum-spherical-polar-expansion}
\begin{split}
  \frac{\partial( \rho v_r)}{\partial t}
  + \frac{1}{r^2}\frac{\partial}{\partial r} \left( r^2 \rho v_r^2 + r^2 \rho c_s^2 \right)\\
  + \frac{1}{r \sin\theta} \frac{\partial}{\partial \theta} \left(\rho v_r v_\theta \sin\theta\right)
  + \frac{1}{r\sin\theta}\frac{\partial}{\partial\phi}\left(\rho v_r v_\phi\right)\\
  - \frac{\rho v_\theta^2 + \rho v_\phi^2 + 2\rho c_s^2}{r} = \rho g_r.    
\end{split}
\end{equation}
Here, $\rho$ is the gas density, $v_r$, $v_\theta$, and $v_\phi$ are the radial, meridional, and azimuthal components of the gas velocity, and $g_r$ is the radial component of the gravitational acceleration.
Integrating \cref{eq:continuity-spherical-polar-expansion,eq:momentum-spherical-polar-expansion} over a full solid angle and dividing by $4\pi$ leads to the angle-averaged continuity and momentum equation,
\begin{equation}\label{eq:continuity-angle-averaged}
  \frac{\partial \left<\rho \right>}{\partial t} + \frac{1}{r^2}\frac{\partial}{\partial r} \left( r^2 \left<\rho v_r \right> \right) = 0,
\end{equation}
\begin{equation}\label{eq:momentum-angle-averaged}
\begin{split}
  \frac{\partial \left<\rho v_r \right>}{\partial t}
  + \frac{1}{r^2} \frac{\partial}{\partial r} \left( r^2 \left<\rho v_r^2 \right> + r^2 \left<\rho c_s^2 \right> \right)\\
  - \frac{\left<\rho v_\theta^2 \right> + \left<\rho v_\phi^2 \right> + 2 \left<\rho c_s^2 \right>}{r}
  = \left<\rho g_r \right>,    
\end{split}
\end{equation}
in which angle brackets denote the averaging operation
\begin{equation}
  \left<Q \right>\equiv \frac{1}{4\pi}\int_0^{2\pi}\int_0^\pi Q \sin\theta d\theta d\phi
\end{equation}
for any physical quantity $Q$.
It is useful to define a related operation
\begin{equation}\label{eq:mass-weighted-average}
  \left<Q \right>_\rho \equiv \frac{\int_0^{2\pi} \int_0^\pi \rho Q \sin\theta d\theta d\phi}{\int_0^{2\pi}\int_0^\pi \rho\sin\theta d\theta d\phi} = \frac{\left<\rho Q\right>}{\left<\rho\right>}
\end{equation}
which is the mass-weighted angle-average of $Q$.

Without loss of generality, we decompose the velocity fields into mean and turbulent components,
\begin{align}
  v_r &= \left<v_r \right>_\rho + \delta v_r,\label{eq:decomposition1}\\
  v_\theta &= \left<v_\theta \right>_\rho + \delta v_\theta,\label{eq:decomposition2}\\
  v_\phi &= \left<v_\phi \right>_\rho + \delta v_\phi.\label{eq:decomposition3},
\end{align}
such that $\left<\delta v_r \right>_\rho = \left<\delta v_\theta\right>_\rho = \left<\delta v_\phi \right>_\rho = 0$ by definition.
Using \cref{eq:continuity-angle-averaged,eq:mass-weighted-average,eq:decomposition1,eq:decomposition2,eq:decomposition3}, one can recast \cref{eq:momentum-angle-averaged} into a Lagrangian equation of motion 
\begin{equation}\label{eq:lagrangian_eom}
  \left<\frac{D \left<v_r \right>_\rho}{Dt} \right>_\rho
  = f_\mathrm{thm} + f_\mathrm{trb} + f_\mathrm{grv} + f_\mathrm{cen} + f_\mathrm{ani}
\end{equation}
where
\begin{equation}
    \left<\frac{D \left<v_r \right>_\rho}{Dt} \right>_\rho
    = \frac{\partial \left<v_r \right>_\rho}{\partial t} + \left<v_r \right>_\rho \frac{\partial \left<v_r \right>_\rho}{\partial r}
\end{equation}
is the radial acceleration.
The individual force components (thermal, turbulent, gravitational, centrifugal, and anisotropic terms) are given by
\begin{align}
  f_\mathrm{thm} &= -\frac{1}{\left<\rho \right>}\frac{\partial P_\mathrm{thm}}{\partial r}\label{eq:fthm},\\
  f_\mathrm{trb} &= -\frac{1}{\left<\rho \right>}\frac{\partial P_\mathrm{trb}}{\partial r}\label{eq:ftrb},\\
  f_\mathrm{grv} &= \left< g_r \right>_\rho,\\
  f_\mathrm{cen} &= \frac{\left<v_\theta \right>_\rho^2 + \left<v_\phi \right>_\rho^2}{r},\\
  f_\mathrm{ani} &= \frac{\left<\delta v_\theta^2 \right>_\rho + \left<\delta v_\phi^2 \right>_\rho - 2 \left<\delta v_r^2 \right>_\rho}{r}\label{eq:fani},
\end{align}
in which the thermal and turbulent pressures are defined by
\begin{align}
  P_\mathrm{thm} &\equiv \left<\rho \right>c_s^2,\\
  P_\mathrm{trb} &\equiv \left<\rho \right>\left<\delta v_r^2 \right>_\rho,
\end{align}
respectively.
In what follows, we assume the centrifugal force is negligible compared to other forces and the turbulence is statistically isotropic, such that $f_\mathrm{cen} = f_\mathrm{ani} = 0$.

\section{Family of Equilibria}\label{sec:equilibrium_sequence}

We now consider a roughly spherical, radially stratified region, and seek a solution in which thermal and turbulent pressure gradient forces balance self-gravity at every radius, i.e.
\begin{equation}\label{eq:force_balance}
  f_\mathrm{thm} + f_\mathrm{trb} = -f_\mathrm{grv}.
\end{equation}
Within a region relatively far from surrounding gravitating masses, the gravitational force can be approximated by
\begin{equation}\label{eq:approx_gravity}
  f_\mathrm{grv} \approx -\frac{GM_\mathrm{enc}(r)}{r^2},
\end{equation}
where
\begin{equation}\label{eq:def_menc}
  M_\mathrm{enc}(r) \equiv 4\pi\int_0^r  r'^2\langle \rho\rangle\,dr'
\end{equation}
is the enclosed mass within the radius $r$.
Using \cref{eq:fthm,eq:ftrb,eq:approx_gravity,eq:def_menc}, \cref{eq:force_balance} can be written as
\begin{equation}\label{eq:dimensional_steady_equilibrium}
  \frac{1}{r^2} \frac{\partial}{\partial r} \left( \frac{r^2}{\left<\rho \right>}\frac{\partial P_\mathrm{eff}}{\partial r}  \right) = -4 \pi G \left<\rho \right>
\end{equation}
where 
\begin{equation}\label{eq:effective-pressure}
  P_\mathrm{eff} \equiv P_\mathrm{thm} + P_\mathrm{trb}
\end{equation}
is the total effective pressure.
Motivated by observations, we further assume that the turbulent velocity dispersion increases with radius as a power law,
\begin{equation}\label{eq:linewidth_size}
  \left<\delta v_r^2 \right>_\rho^{1/2} = c_s \left( \frac{r}{r_s} \right)^{p},
\end{equation}
in which $r_s$ is the sonic radius and $p$ is the power law index, such that $P_\mathrm{trb} = P_\mathrm{thm}(r/r_s)^{2p}$.

To recast \cref{eq:dimensional_steady_equilibrium} in a dimensionless form, we define a dimensionless radial coordinate $\xi$ and logarithmic density contrast $u$ by\footnote{See Appendix \ref{app:alt_formulation} for an alternative dimensionless formulation based on given external pressure.}
\begin{align}
  \xi &\equiv \frac{\left(4\pi G\rho_c\right)^{1/2}}{c_s}r,\label{eq:def_xi}\\
  u(\xi;\xi_s,p) &\equiv \ln \frac{\rho_c}{\left<\rho \right>}\label{eq:def_u},
\end{align}
where $\rho_c$ is the density at the center (i.e., $r=0$).
In \cref{eq:def_u}, it is made explicit that $u$ is a function of $\xi$ and involves two dimensionless parameters $\xi_s = r_s(4\pi G \rho_c)^{1/2}/c_s$ and $p$.
\REV{Here, $\xi_s$ is the dimensionless sonic radius defined by $r=r_s$ in \cref{eq:def_xi}.}
We also define the dimensionless enclosed mass
\begin{equation}\label{eq:enclosed_mass}
  m(\xi;\xi_s,p) = \frac{(4\pi G^3\rho_c)^{1/2}}{c_s^3}M_\mathrm{enc}
\end{equation}
and the function
\begin{equation}\label{eq:def_f}
  \chi(\xi;\xi_s,p) \equiv 1 + \left(\frac{\xi}{\xi_s}\right)^{2p}
\end{equation}
that relates thermal pressure to total pressure through $P_\mathrm{eff} = \chi P_\mathrm{thm}$.
In terms of these dimensionless variables, \cref{eq:dimensional_steady_equilibrium} becomes
\begin{equation}\label{eq:emden}
  \frac{1}{\xi^2}\frac{\partial}{\partial \xi} \left[ \xi^2 \left(\chi \frac{\partial u}{\partial \xi} - \frac{\partial \chi}{\partial \xi}\right) \right] = e^{-u}.
\end{equation}
It is useful to note that the term in the square bracket is identical to the dimensionless enclosed mass, i.e.,
\begin{equation}\label{eq:def_m}
  m(\xi;\xi_s,p) \equiv \xi^2 \left(\chi\frac{\partial u}{\partial \xi} - \frac{\partial \chi}{\partial \xi} \right).
\end{equation}
We remind the reader that \cref{eq:emden} reduces to the usual isothermal Lane-Emden equation \citep{chandrasekhar49} in the limit of vanishing turbulent pressure.

For a solution to have a finite central density (i.e., a plateau), it must satisfy the boundary conditions
\begin{align}
  u\rvert_{\xi=0} &= 0,\\
  \frac{\partial u}{\partial \xi}\bigg\rvert_{\xi=0} &= u'_0,
\end{align}
with a finite value of $u'_0$ which does not necessarily equal zero.
For example, for $p=0.5$, one can find a series solution of \cref{eq:emden} near $\xi=0$,
\begin{equation}
  u = \frac{1}{\xi_s}\xi + \left( \frac{1}{6} - \frac{1}{2\xi_s^2} \right) \xi^2 + \left( \frac{1}{3\xi_s^3} - \frac{7}{36\xi_s} \right) \xi^3 + \cdots,
\end{equation}
which yields $u'_0 = \xi_s^{-1}$.

\REV{In practice, one can integrate \cref{eq:emden} in terms of the logarithmic radius $\varpi \equiv \ln \xi$, with the initial conditions $u=0$ and $\partial u/\partial \varpi=\xi (\partial u/\partial\xi) =0$ for very small $\xi$.}
The resulting solutions form a two-parameter family characterized by $\xi_s$ and $p$.
We term this two-parameter family of solutions \acfp{TES}.
\REV{A python package that calculates these solutions is available in \citet{tesphere}.}
\footnote{\url{https://github.com/sanghyukmoon/turbulent_equilibrium_sphere}.}

\begin{figure}[htpb]
  \plotone{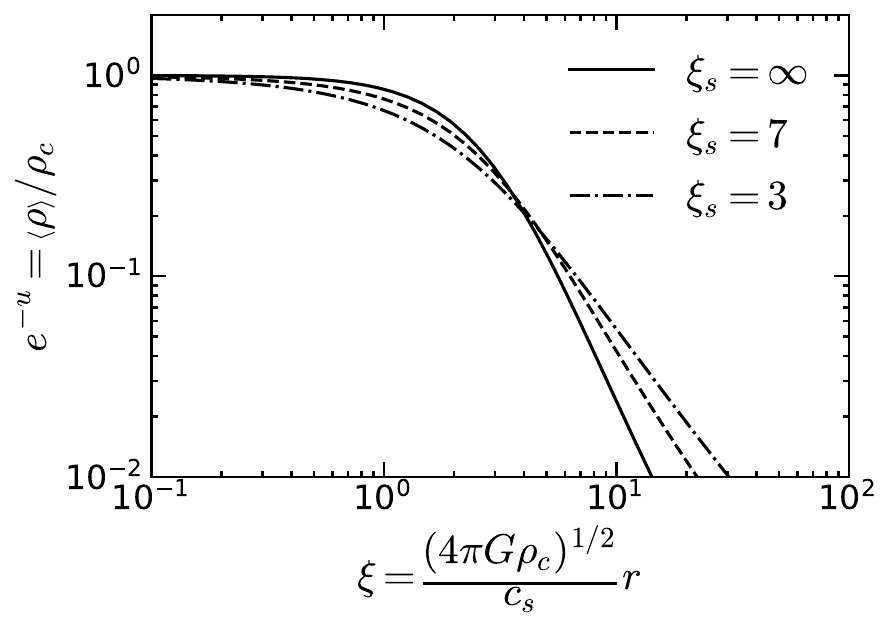}
  \caption{Density profiles of TES with linewidth-size slope $p=0.5$, for selected dimensionless sonic radii $\xi_s = \infty$ (solid), $7$ (dashed), and $3$ (dot-dashed).
  The profile with $\xi_s=\infty$ is identical to the \ac{BE} solution, while those with smaller $\xi_s$ correspond to turbulent equilibria (higher amplitude turbulence for smaller $\xi_s$).
  The density is normalized with respect to the central density $\rho_c$.}
  \label{fig:tes_density_profiles}
\end{figure}

\cref{fig:tes_density_profiles} plots the radial density profiles of the \acp{TES} having $p=0.5$ and $\xi_s=\infty$, $7$, and $3$.
The profile with $\xi_s=\infty$ is identical to that of the \ac{BE} sphere.
\REV{For smaller values of $\xi_s$, the density profile in the inner part is steeper than the \ac{BE} solution, while it is shallower in the outer part. 
In the innermost regions of cores where gravity is weak and the  local density profile $\rho\propto r^{-q}$ is shallow (small $q$), thermal pressure gradients for the \ac{TES} must be steeper than for the \ac{BE} sphere in order to compensate for turbulent pressure forces, which are {\it inward} when $q < 2p$.
Conversely, in the outer regions of cores where the \ac{BE} solution requires a steep density gradient for thermal pressure to balance gravity, the density gradient can be shallower for the \ac{TES} because turbulent pressure support assists thermal pressure support provided $q > 2p$.
}

\begin{figure}[htpb]
  \plotone{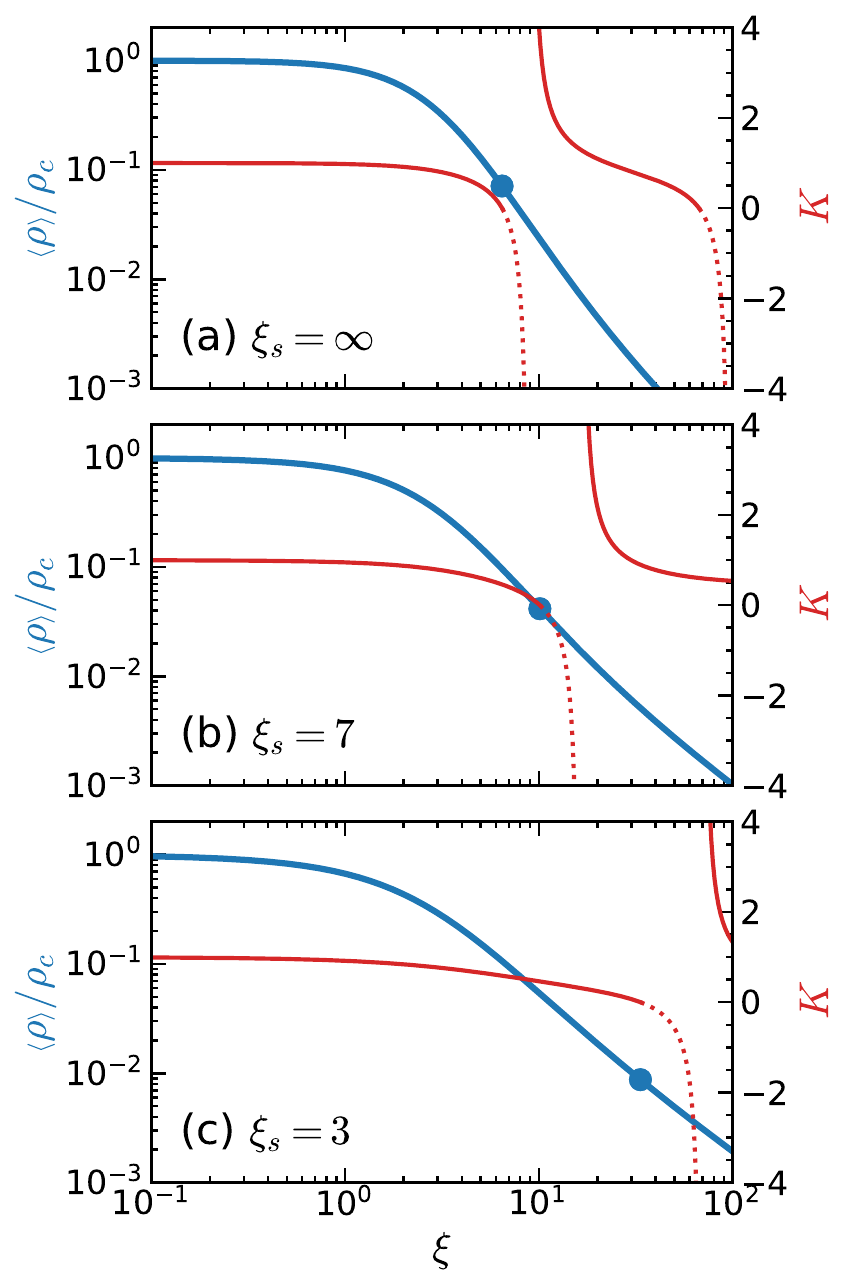}
  \caption{Radial profiles of the angle-averaged density (blue solid line, left axis) for TES solutions with (a) $\xi_s = \infty$, i.e. zero turbulence, (b) $\xi_s = 7$, i.e. low turbulence, and (c) $\xi_s = 3$, i.e. higher turbulence.
  For all three panels, $p = 0.5$.
  Also shown is the dimensionless bulk modulus $K$ (\cref{eq:bulk_modulus}, red line, right axis); positive values are shown solid and negative dotted.
  In each panel, the point at which $K$ first becomes negative is marked with a blue circle.
  This is the critical solution; solutions with smaller radii are stable and those with larger radii are unstable. 
  }
  \label{fig:bulk_modulus}
\end{figure}
\section{Stability}\label{sec:stability}

We assess the stability of the \ac{TES} using an analysis similar to \citet{bonnor56}.
We consider a sphere of radius $r$ and imagine that the volume enclosed is slightly compressed.
In reality, this compression can be driven by either inflows from larger radii or random turbulent motions.
As the gas distribution interior to this Lagrangian boundary surface adjusts to a new equilibrium, the effective pressure at the boundary will change according to the normalized bulk modulus defined by
\begin{equation}\label{eq:bulk_modulus}
  K \equiv -\left(\frac{\partial \ln P_\mathrm{eff}}{\partial \ln V_r}\right)_{M_\mathrm{enc},c_s,r_s,p}
\end{equation}
where
\begin{equation}\label{eq:enclosed-volume}
  V_r \equiv \frac{4\pi r^3}{3}
\end{equation}
is the spherical volume enclosed within the radius $r$.
Note that we keep the sonic radius $r_s$ and index $p$ constant in \cref{eq:bulk_modulus}, assuming that the background turbulent flow is not affected by the perturbations applied.\footnote{This implies that the turbulent kinetic energy contained within the Lagrangian radius (i.e., the radius containing constant $M_\mathrm{enc}$) decreases under the compression. Although turbulence can in principle be amplified when the compression timescale is shorter than the flow crossing time \citep{robertson12}, in practice we do not expect that the overall compression speed of turbulent cores will exceed their velocity dispersion, unless they are already undergoing gravitational collapse.}

\begin{figure*}[htpb]
  \plotone{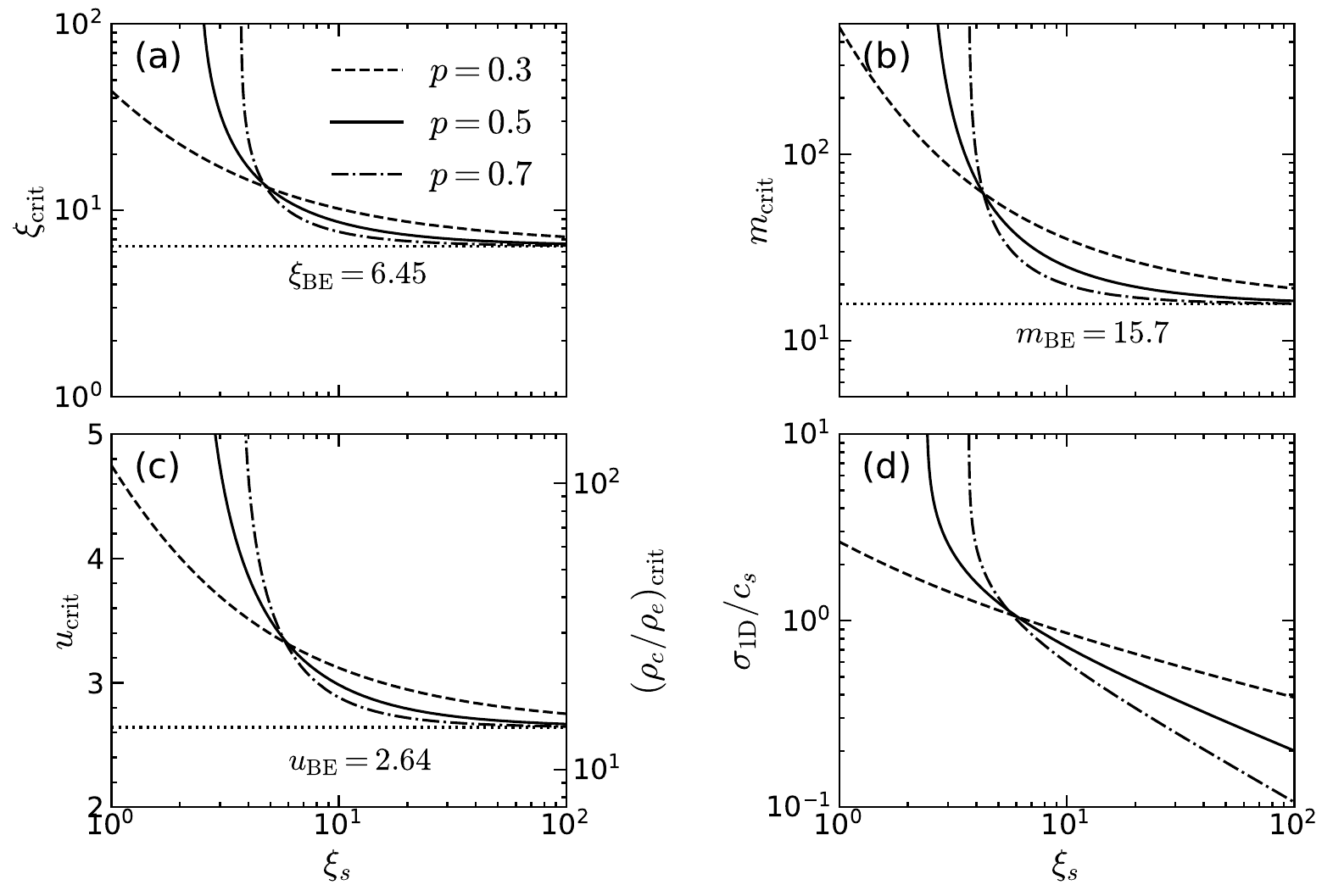}
  \caption{Parameteric dependence of the critical \ac{TES} properties on the dimensionless sonic radius $\xi_s$, for $p=0.3$ (dashed), $0.5$ (solid), and $0.7$ (dot-dashed).
  (a) the dimensionless critical radius $\xi_\mathrm{crit}$
  (b) the dimensionless critical mass $m_\mathrm{crit}$
  (c) the critical density contrast $u_\mathrm{crit}$ (left axis) and $(\rho_c/\rho_e)_\mathrm{crit}\equiv \exp(u_\mathrm{crit})$ (right axis).
  (d) the mass-weighted average turbulent Mach number $\sigma_\mathrm{1D}/c_s$.
  In the panels (a)--(c), the corresponding critical values for the \ac{BE} limit (i.e., $\xi_s\to \infty$) is plotted in dotted lines.}
  \label{fig:critical_tes}
\end{figure*}

To calculate the bulk modulus $K$, we first write the effective pressure, volume, and mass as functions of $\xi$ and $\xi_s$ as follows:
\begin{align}
  P_\mathrm{eff} &=\rho_c c_s^2 \chi e^{-u}= \frac{c_s^4}{4\pi G r_s^2}\xi_s^2\chi e^{-u}\label{eq:peff},\\
  V_r &= \frac{4\pi}{3}r^3=\frac{4\pi r_s^3}{3} \xi_s^{-3}\xi^3 \label{eq:Vr},\\
  M_\mathrm{enc} &=\frac{c_s^3}{(4\pi G^3 \rho_c)^{1/2}}m= \frac{c_s^2 r_s}{G} \xi_s^{-1} m \label{eq:Menc}.
\end{align}
We remind the reader that $\chi$, $u$, and $m$ appearing in \cref{eq:peff,eq:Vr,eq:Menc} are dimensionless functions of the variable $\xi$ and the parameters $\xi_s$ and $p$.

The total derivatives of $\ln P_\mathrm{eff}$, $\ln V_r$, and $\ln M_\mathrm{enc}$ at a fixed $c_s$, $r_s$, and $p$ are
\begin{align}
  \begin{split}\label{eq:dlnPeff}
    \delta \ln P_\mathrm{eff} &= \left( \frac{\partial\ln \chi}{\partial\ln\xi} - \frac{\partial u}{\partial \ln \xi} \right) \delta\ln\xi\\
                              &\quad + \left( 2+ \frac{\partial\ln \chi}{\partial\ln\xi_s} - \frac{\partial u}{\partial\ln\xi_s} \right) \delta\ln\xi_s\\
                              &= -\frac{m}{\chi\xi} \delta\ln\xi\\
                              &\quad + \left(2+ \frac{\partial\ln \chi}{\partial\ln\xi_s}  - \frac{\partial u}{\partial\ln\xi_s} \right) \delta\ln\xi_s,
  \end{split}\\
  \delta \ln V_r &= 3\,\delta\ln\xi - 3\,\delta\ln\xi_s\label{eq:dlnVr},\\
  \delta \ln M_\mathrm{enc} &= \frac{\partial \ln m}{\partial \ln \xi}\delta\ln \xi + \left( \frac{\partial \ln m}{\partial \ln \xi_s} - 1 \right) \delta \ln \xi_s\label{eq:dlnMenc},
\end{align}
where we have used \cref{eq:def_m} in the second equality of \cref{eq:dlnPeff}.
Because we are interested in the bulk modulus of a region of constant $M_\mathrm{enc}$, \cref{eq:dlnMenc} requires
\begin{equation}\label{eq:differential}
  \frac{\delta \ln \xi}{\delta \ln \xi_s} = \frac{m}{\xi^3 e^{-u}} \left( 1 - \frac{\partial \ln m}{\partial \ln \xi_s} \right)
\end{equation}
to be satisfied.
Using \cref{eq:dlnPeff,eq:dlnVr,eq:differential}, after some algebra, one obtains
\begin{align}
  K &= \frac{2}{3} \frac{1 - \frac{1}{2} \frac{\partial u}{\partial\ln\xi_s} + \frac{1}{2}\frac{\partial\ln \chi}{\partial\ln\xi_s} - \frac{m^2}{2\chi e^{-u}\xi^4} \left( 1 - \frac{\partial\ln m}{\partial\ln\xi_s} \right) }{1 - \frac{m}{\xi^3e^{-u}} \left( 1 - \frac{\partial\ln m}{\partial\ln\xi_s} \right) }\label{eq:bulk_modulus_dimensionless}\\
    &= \frac{2}{3}\frac{1 - \frac{1}{2}\frac{\partial u}{\partial \ln \xi_s} + \frac{1}{2}\frac{\partial \ln \chi}{\partial \ln \xi_s} - \left( \frac{4\pi}{3} \right)^{1/3} \frac{GM_\mathrm{enc}^2}{6P_\mathrm{eff}V_r^{4/3}}\left( 1 - \frac{\partial \ln m}{\partial \ln \xi_s} \right) }{1 - \frac{M_\mathrm{enc}c_s^2}{3P_\mathrm{thm}V_r}\left( 1 - \frac{\partial \ln m}{\partial \ln \xi_s} \right)}.\label{eq:bulk_modulus_dimensional}
\end{align}
In the limit of negligible turbulence ($\xi_s\to \infty$), each of the derivatives with respect to $\xi_s$ in \cref{eq:bulk_modulus_dimensional} may be set to zero, and we recover Equation 2.16 of \citet{bonnor56} for an isothermal sphere.
In the simultaneous limit $\xi_s\to \infty$ and $G\to 0$, the density is uniform so that $M_\mathrm{enc}c_s^2/(P_\mathrm{thm}V_r) = 1$, and one recovers the ideal gas $K=1$.
The equilibrium is unstable when $K < 0$, because a slight compression leads to a further decrease in the interior pressure.

\cref{fig:bulk_modulus} plots density profiles for selected \acp{TES} with $p=0.5$ and $\xi_s=\infty$, $7$, and $3$, together with the radial profile of $K$ in each case.
In all cases, for $\xi\ll 1$ the solutions have $K\approx 1$ because there is not enough mass to be self-gravitating and the thermal pressure dominates the turbulent pressure.
As $\xi$ increases, however, self-gravity becomes more and more important and $K$ decreases.
We define the critical radius $\xi_\mathrm{crit}$ as the radius where $K$ first becomes negative. 
For $\xi>\xi_\mathrm{crit}$, the magnitude of $K$ rapidly increases, undergoing another sign change at $\xi = \xi_\mathrm{crit,2}$ back to positive.
A \ac{TES} with outer radius $\xi > \xi_\mathrm{crit}$ is unstable because a part of its interior has $K < 0$.

For each member of the family of solutions with given $p$ and $\xi_s$, the \ac{TES} with outer radius $\xi = \xi_\mathrm{crit}$ is identified as the critical solution.  
We further define the critical logarithmic density contrast $u_\mathrm{crit}\equiv u(\xi_\mathrm{crit})=\ln[\rho_c/\rho(\xi_\mathrm{crit})]$ and the critical mass $m_\mathrm{crit}\equiv m(\xi_\mathrm{crit})$ associated with the critical radius $\xi_\mathrm{crit}$.
\REV{\cref{fig:critical_tes}(a)--(d) shows the parametric dependence of $\xi_\mathrm{crit}$, $m_\mathrm{crit}$, $u_\mathrm{crit}$, and the average turbulent velocity dispersion $\sigma_\mathrm{1D}$ on $\xi_s$, for $p=0.3$, $0.5$, and $0.7$.  
Here, we define $\sigma_\mathrm{1D}$ using the mass-weighted average over the whole core within $r_\mathrm{crit}$, assuming the turbulence is statistically isotropic:
}
\begin{equation}\label{eq:sigma_r}
\begin{split}
  \sigma_\mathrm{1D} &\equiv \left(\frac{\iiint_{r<r_\mathrm{crit}} \rho \delta v_r^2 dV}{\iiint_{r<r_\mathrm{crit}} \rho dV}\right)^{1/2}\\
                     &= c_s \left[\frac{\int_0^{\xi_\mathrm{crit}} e^{-u} (\xi/\xi_s)^{2p} \xi^2 d\xi}{\int_0^{\xi_\mathrm{crit}} e^{-u} \xi^2 d\xi}\right]^{1/2}.    
\end{split}
\end{equation}

In the limit of $\xi_s\to \infty$, we find $\xi_\mathrm{crit}=6.45$, $m_\mathrm{crit}=15.7$, and $u_\mathrm{crit}=2.64$, identical to the well-known result for the critical \ac{BE} sphere \citep{bonnor56,lynden-bell68,sormani13}.

\REV{The point $\xi_s = \xi_\mathrm{crit}$ might be of some interest, because the sonic radius is within the core for smaller $\xi_s$, while it is outside the core for larger $\xi_s$.
For $p=0.3$, $0.5$, and $0.7$, $\xi_s=\xi_\mathrm{crit}$ occurs at $\xi_s = 10.2$, $8.99$, and $8.21$.
It is interesting to note that $\sigma_\mathrm{1D} \approx c_s$ at $\xi_s \approx 6$, quite insensitive to $p$.
It is worth mentioning, as well, that while this and other curves plotted in \cref{fig:critical_tes} give an impression that they all cross at the same locus, a detailed examination reveals that it is only approximately true, and the crossing points differ for different curves (e.g. $\xi_\mathrm{crit} \approx 14$ when $\xi_s \approx 5$, and $m_\mathrm{crit}\approx 65$ when $\xi_s \approx 4$).
The similar crossing points arise because the curves with different $p$ (for panels a-c) go to the same limit $\xi_\mathrm{crit}\to\xi_\mathrm{BE}$ at $\xi_s\to\infty$, and more generally these solutions must be quite similar in the regime where $\xi_s>\xi_\mathrm{crit}$ because $\chi$ in \cref{eq:def_f} is close to unity.}
For $p=0.5$, \cref{tb:tes} provides tabulated values for $\xi_s$, $\xi_\mathrm{crit}$, $m_\mathrm{crit}$, $u_\mathrm{crit}$, and $\sigma_\mathrm{1D}$.
\begin{deluxetable}{ccccc}
  \tablecaption{Properties of critical \acp{TES} with $p=0.5$.
  \label{tb:tes}}
  \tablehead{
\colhead{$\,\, \xi_s             \,\,$} &
\colhead{$\,\, \xi_\mathrm{crit} \,\,$} &
\colhead{$\,\, m_\mathrm{crit}   \,\,$} &
\colhead{$\,\, u_\mathrm{crit}   \,\,$} &
\colhead{$\,\, \sigma_\mathrm{1D}   \,\,$}
  }
  \startdata
  $\infty$ & 6.45 & 15.7 & 2.64 & 0.0 \\
  18.2 & 7.50 & 19.9 & 2.81 & 0.5 \\
  6.42 & 10.6 & 34.5 & 3.25 & 1.0 \\
  4.21 & 15.8 & 64.3 & 3.76 & 1.5 \\
  3.42 & 22.7 & 115 & 4.24 & 2.0 \\
  2.55 & 103 & $1.4\times 10^3$ & 6.17 & 5.0 \\
  2.44 & 385 & $1.3\times 10^4$ & 7.79 & 10 \\
  2.42 & $\infty$ & $\infty$ & $\infty$ & $\infty$
  \enddata
  \tablecomments{\cref{tb:tes} is published in its entirety in machine-readable format. A few representative rows are shown here for guidance regarding its form and content.}
\end{deluxetable}

We note that $\xi_\mathrm{crit}$, $m_\mathrm{crit}$, and $u_\mathrm{crit}$ can be translated to the dimensional critical radius $r_\mathrm{crit}$, mass $M_\mathrm{crit}$, and critical density contrast $(\rho_c / \rho_e)_\mathrm{crit}$,  through \cref{eq:def_xi,eq:def_u,eq:enclosed_mass}.
For radius and mass, conversion between dimensionless and dimensional variables is given by:
\begin{equation}\label{eq:r_xi}
  r =  \xi \times 0.014 \,\pc\left( \frac{T}{10\,\mathrm{K}} \right)^{1/2} \left( \frac{{n}_\mathrm{H,c}}{10^5\,\mathrm{cm}^{-3}} \right)^{-1/2} 
\end{equation}
\begin{equation}\label{eq:M_m}
  M =  m\times 0.11 \,M_\odot \left( \frac{T}{10\,\mathrm{K}} \right)^{3/2} \left( \frac{{n}_\mathrm{H,c}}{10^5\,\mathrm{cm}^{-3}} \right)^{-1/2} 
\end{equation}
where ${n}_\mathrm{H,c}$ is the central density and $T$ is the temperature.

\cref{fig:critical_tes} shows that the dimensionless critical radius $\xi_\mathrm{crit}$ steeply increases with decreasing $\xi_s$.
\REV{This indicates that a core forming in a region of strong local turbulence would initially be stable: large $\xi_\mathrm{crit}$ would make the critical radius lie beyond the effective tidal radius that limits the core, as imposed by the gravity of nearby structures in the \ac{GMC}
However, evolution generally occurs in the direction of increasing $\xi_s \propto \rho_c^{1/2} r_s$ as the initial converging flows further compress the core and turbulence dissipates, leading to a decrease in $\xi_\mathrm{crit}$.
When $\xi_\mathrm{crit}$ becomes small enough, the core will become unstable, triggering collapse.
More details on the collapse scenario will be discussed in \cref{sec:collapse_scenario}; the quantitative test of this scenario using numerical simulations will be presented in Paper II.}

We empirically find that, for $p \gtrsim 0.4$, there exists a minimum dimensionless sonic radius $\xi_{s,\mathrm{min}}$ below which $K>0$ at every radius.
In \cref{fig:critical_tes}, $\xi_{s,\mathrm{min}}$ for a given $p$ corresponds to the value at which $\xi_\mathrm{crit}\rightarrow \infty$. 
The existence of $\xi_{s,\mathrm{min}}$ for a \ac{TES} family means that for a given central density, every member of the family is {\it stable} if the turbulence is sufficiently strong that $r_s < \xi_{s,\mathrm{min}} c_s/ (4\pi G \rho_c)^{1/2}$, where \cref{eq:r_xi} can be used to translate this into physical terms.

Alternatively, because $\xi_s \propto \rho_c^{1/2} r_s$ (\cref{eq:def_xi}), the existence of $\xi_{s,\mathrm{min}}$ implies that for a given \emph{dimensional} sonic radius $r_s$, a \ac{TES} with the central density lower than $\rho_{c,\mathrm{min}} \equiv \xi_{s,\mathrm{min}}^2c_s^2/(4\pi G r_s^2)$ is \emph{always} stable.
Let us suppose that the larger-scale environment of the cores is a spherical cloud of radius $R_\mathrm{cloud}$, mass $M_{\rm cloud}$, and \REV{one-dimensional Mach number $\mathcal{M}_\mathrm{1D} = (R_{\rm cloud}/r_s)^p$}, with virial parameter
\begin{equation}
  \alpha_\mathrm{vir,cloud} \equiv \frac{5 \mathcal{M}_\mathrm{1D}^2c_s^2 R_{\rm cloud}}{G M_{\rm cloud}}.
\end{equation}
The minimum central density for instability to be possible can then be written as
\begin{equation}\label{eq:rho_min}
  \frac{\rho_{c,\mathrm{min}}}{\rho_0} = \frac{\xi_{s,\mathrm{min}}^2}{15} \alpha_\mathrm{vir,cloud} \mathcal{M}_\mathrm{1D}^{2/p - 2},
\end{equation}
where $\rho_0 = M / (4\pi R^3/3)$ is the mean density of the cloud.

For turbulent power law index $p=0.5$, we find $\xi_{s,\mathrm{min}} = 2.42$, which would correspond to a minimum sonic scale of $r_s\approx 0.03 \pc$ using the fiducial central density and temperature in \cref{eq:r_xi}.
Alternatively, the minimum molecular hydrogen number density at the core center, $n_{\mathrm{H2},c}^\mathrm{min}=n_{\rm H,c}^{\rm min}/2 \equiv \rho_{c,\mathrm{min}} / (2\mu_\mathrm{H}m_\mathrm{H})$, that would allow for instability is given in terms of the physical sonic scale $r_s$ by
\begin{equation}\label{eq:nc_min}
  n_{\mathrm{H2},c}^\mathrm{min} = 2.2\times 10^4\,\mathrm{cm}^{-3}\,\left(\frac{\xi_{s,\mathrm{min}}}{2.42}\right)^2\left( \frac{T}{10\,\mathrm{K}} \right) \left( \frac{r_s}{0.05\,\mathrm{pc}} \right)^{-2}.
\end{equation}
\REV{We emphasize that these are the \emph{local} values of $T$ and $r_s$, the latter of which can vary considerably within a \ac{GMC} (see Paper II)}.
With typical molecular cloud virial parameter $\alpha_\mathrm{vir,cloud}\sim 2-4$ on $\sim 100 \pc$ scale \citep{Sun2020},  \cref{eq:rho_min} implies that the central density in the core would have to exceed the ambient cloud density by at least a factor $\rho_{c,\mathrm{min}}/\rho_0\sim {\cal M}^2$ for instability to be possible, \REV{when $p=0.5$}.

We stress that the condition $\xi_s> \xi_{s,{\rm min}}$ is a necessary but not a sufficient condition for collapse.
That is, $\rho_c > \rho_{c,\mathrm{min}}$  does \emph{not} guarantee instability and hence should not be interpreted as a critical density for collapse.
Instead, $\rho_c > \rho_{c,\mathrm{min}}$ simply means that there exists a critical radius beyond which a quasi-equilibrium is unstable.
For a core to collapse, its central density must exceed the minimum value, \emph{and} the core's outer radius and total mass must exceed the critical values for \ac{TES} solutions, as shown in \cref{fig:critical_tes}.

\begin{figure}[htpb]
  \plotone{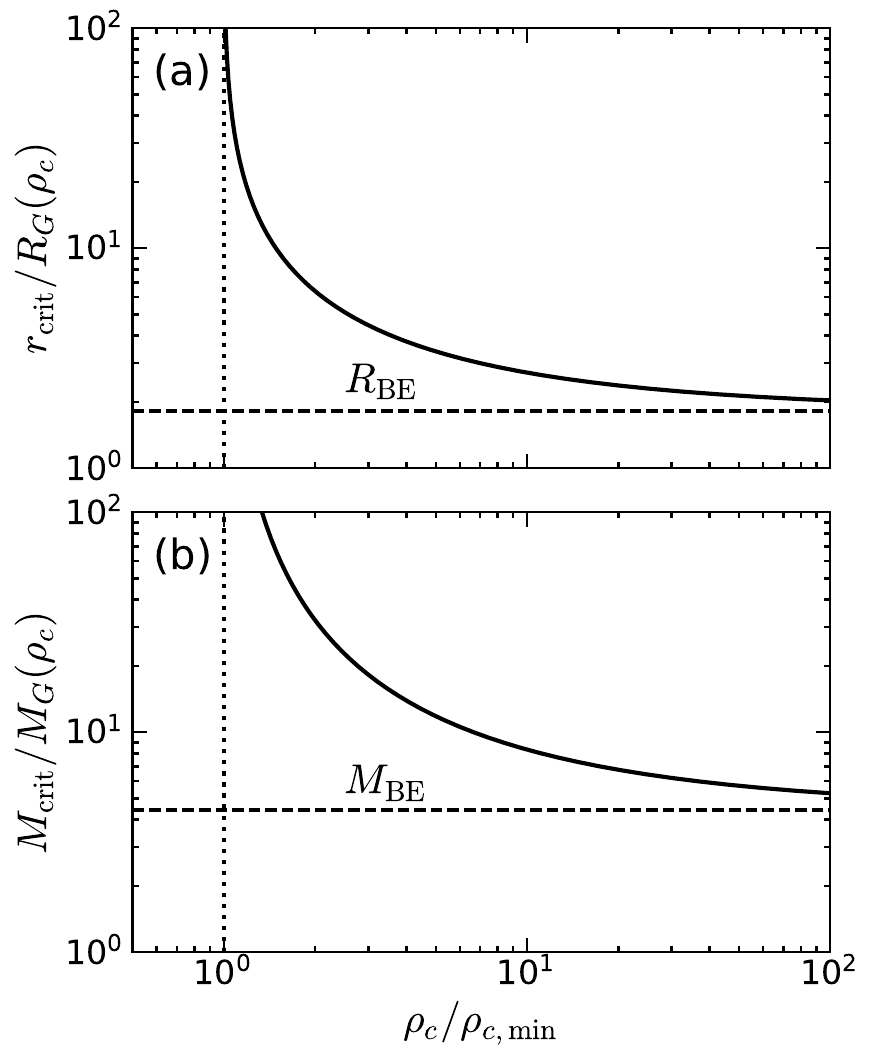}
  \caption{(a) The ratio of the critical radius to the gravitational length scale $R_G = c_s / (G^{1/2}\rho_c^{1/2})$ (see \cref{eq:rgrav} for dimensional conversion), as a function of $\rho_c/\rho_{c,\mathrm{min}}=(\xi_s/\xi_{s,\mathrm{min}})^2$, for $\ac{TES}$ with turbulence index $p=0.5$.
  (b) A similar ratio for the critical mass and the gravitational mass scale $M_G = c_s^3 / (G^{3/2}\rho_c^{1/2})$.
  The corresponding ratios for the critical \ac{BE} sphere are shown as dashed lines in each panel.
  A core with $\rho_c/\rho_{c,\mathrm{min}}<1$ cannot collapse because the critical radius and mass for instability is undefined.
  When $\rho_c/\rho_{c,\mathrm{min}}>1$, a core \emph{may} collapse, provided its outer radius and mass exceed $r_\mathrm{crit}$ and $M_\mathrm{crit}$, respectively, as shown here.}
  \label{fig:minimum_density}
\end{figure}

For $p=0.5$, \cref{fig:minimum_density} plots the critical radius and mass as a function of $\rho_c / \rho_{c,\mathrm{min}}=(\xi_s/\xi_{s,\mathrm{min}})^2$.
We note that cases with large (small) $\rho_c/\rho_{c,\mathrm{min}}$ correspond to weak (strong) turbulence. 
A highly turbulent core would therefore have to be larger and more massive to become unstable, compared to quiescent cores.
Notably, \cref{fig:minimum_density} shows that when $\rho_c$ is close to $\rho_{c,\mathrm{min}}$, $r_\mathrm{crit}$ is at least a factor of a few larger than the \ac{BE} radius at that $\rho_c$, $R_\mathrm{BE}=1.82 R_G(\rho_c)$.
\cref{fig:minimum_density} also shows that the mass begins to exceed $M_\mathrm{BE} = 4.43 M_G(\rho_c)$ by a factor two or more when $\rho_c/\rho_{c,\mathrm{min}} < 9$.
\REV{At the same time, \cref{fig:minimum_density} shows that for a given $r_s$ (and therefore given $\rho_{c,\mathrm{min}}$), the central density must become quite large in order for the unstable core mass to approach $M_\mathrm{BE}$: $\rho_c/\rho_{c,\mathrm{min}}$ has to be at least $21$ in order to have a core with $M_\mathrm{crit}/M_\mathrm{BE} < 1.5$.
Since real cores cannot have either too large a radius (this would exceed the cloud size) or too high a central density (difficult to reach with realistic larger-scale dynamics), the majority of cores that become unstable may have a moderate range of $r_\mathrm{crit}/R_\mathrm{BE}$ and $M_\mathrm{crit}/M_\mathrm{BE}$, or equivalently, of $\sigma_\mathrm{1D}/c_s$ (see \cref{fig:critical_tes}).
A comparison of gravitational and flow crossing timescales also suggests a highly turbulent core with $\sigma_\mathrm{1D}\gg c_s$ would be difficult to collapse as a whole (see below).
Although $\xi_{s,\mathrm{min}}$ is a strict lower limit for allowing instability, we expect collapse would mostly occur at $\xi_s \gtrsim 6$, for which turbulent velocity dispersion is at most transonic regardless of the power law index $p$ (see \cref{fig:critical_tes}(d)).}
We further discuss scenarios for collapse in \cref{sec:collapse_scenario}.

\section{Physical Properties of Critical TES}\label{sec:physical_conditions}

\begin{figure*}[htpb]
  \plotone{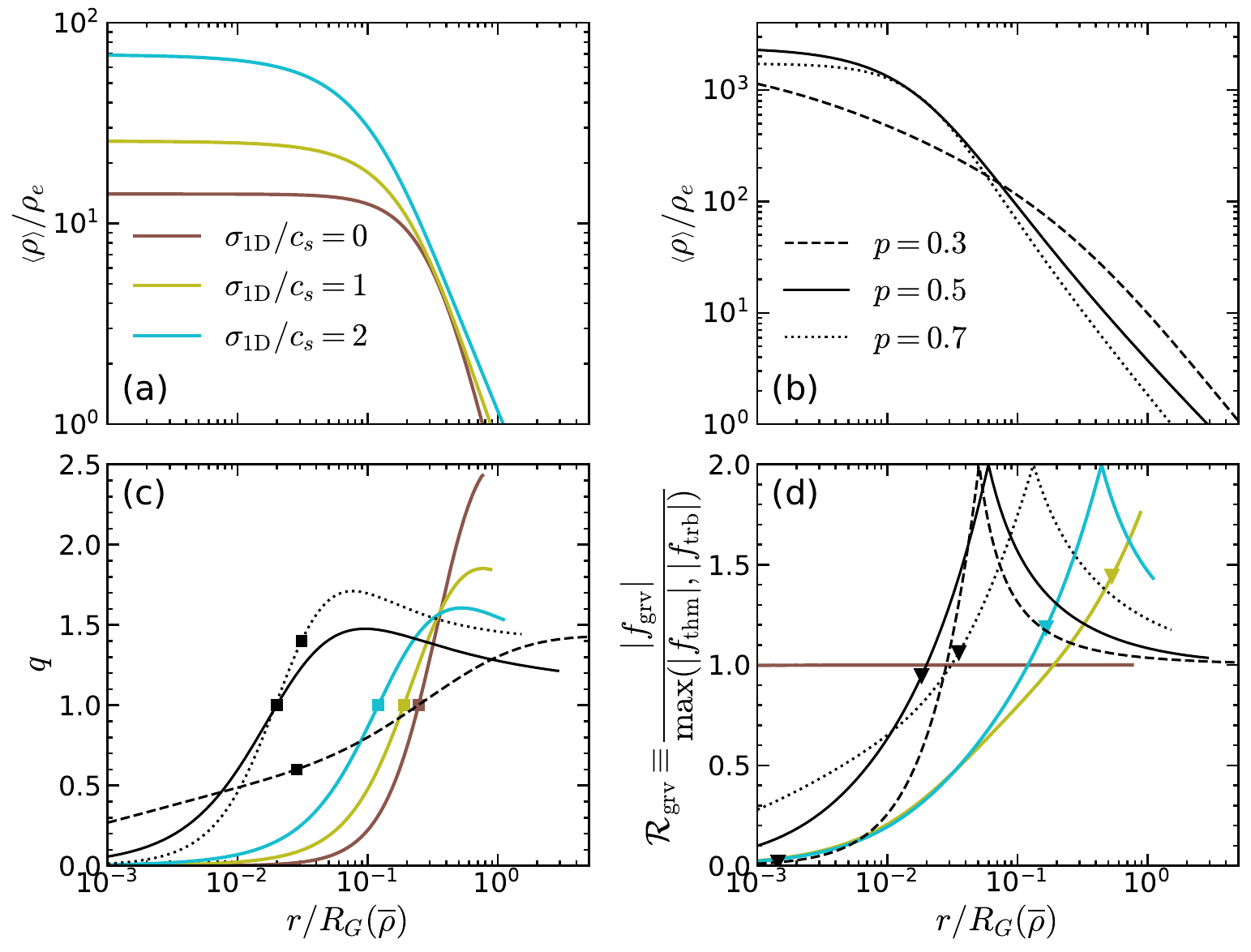}
  \caption{\REV{(a) Density profiles of the critical \acp{TES} with $p=0.5$, for selected mass-weighted average velocity dispersions of $\sigma_\mathrm{1D} = 0$ (brown; equivalent to the critical \ac{BE} sphere), 
  $c_s$ (yellow), and $2c_s$ (cyan).
  The density is normalized to the density at the edge $\rho_e = \left<\rho \right>(r = r_\mathrm{crit})$, and the radius is normalized to the gravitational radius defined in \cref{eq:rgrav}, at the mean core density $\overline{\rho} = M_\mathrm{crit} / (4\pi r_\mathrm{crit}^3/3)$.
  (b) Density profiles of highly turbulent \acp{TES} with $\sigma_\mathrm{1D} = 10c_s$, for selected linewidth-size index $p=0.3$ (dashed), $0.5$ (solid), and $0.7$ (dotted).
  (c) The local power-law slope $q \equiv -\partial \ln \rho / \partial \ln r$ for the curves shown in panels (a) and (b).
  The squares mark the point where $q = 2p$ for each curve: the turbulent pressure gradient force is compressive (supportive) on the left (right) side of this marker.
  (d) The relative strength of the gravitational force compared to the thermal or turbulent pressure gradient force (the larger of the two).
  The maximum value $\mathcal{R} = 2$ is reached when $f_\mathrm{thm} = f_\mathrm{trb}$.
  On the left and right sides of this peak, $|f_\mathrm{thm}| > |f_\mathrm{trb}|$ and $|f_\mathrm{thm}| < |f_\mathrm{trb}|$, respectively.
  The triangles mark the location of the sonic radius.
  }}
  \label{fig:profiles_norm}
\end{figure*}

While the structure of the \ac{TES} is fully characterized by $\xi_s$ and $p$, it is often useful to reparameterize the results in terms of the \REV{core-averaged} one-dimensional velocity dispersion $\sigma_\mathrm{1D}$ \REV{defined in \cref{eq:sigma_r}.}
Here we will consider just the case of critical cores.
For $p=0.5$, \cref{fig:profiles_norm}(a) plots the density profiles of critical \acp{TES} having different values of $\sigma_\mathrm{1D}/c_s$, normalizing relative to the density at the outer radius.
Increased turbulent support makes the core larger and more centrally concentrated; a trans-sonic core has a factor of two higher density contrast than a purely thermal core, while increasing the internal Mach number to 2 leads to a factor $\sim 5$ increase in the density contrast. \REV{Here we show just cases with $p=0.5$, but other indices produce similar profiles for low $\sigma_\mathrm{1D}$ since thermal pressure dominates.}

\REV{\cref{fig:profiles_norm}(a) suggests that a \ac{TES} develops a power law envelope as $\sigma_\mathrm{1D}$ increases.
To illustrate this more clearly, in \cref{fig:profiles_norm}(b) we plot the density profiles of highly turbulent \acp{TES} with $\sigma_\mathrm{1D} = 10 c_s$ for selected linewidth-size indices $p=0.3$, $p=0.5$, and $p=0.7$.
We then calculate the local density slope $q \equiv -\partial \ln\rho/\partial \ln r$ for all cases shown in \cref{fig:profiles_norm}(a)-(b) and plot the profiles of $q$ in \cref{fig:profiles_norm}(c).
While no density profile is described by a single power law, the outer density slopes generally lie in the typical range $q \sim 1\text{--}2$ found in observations \citep{vandertak00,bergin07,chen19}.}

\REV{\cref{fig:profiles_norm}(c) suggests that for $\sigma_\mathrm{1D}/c_s\gtrsim 2$, $q$ becomes flat in the outer part of the core.
In fact, if in the limit $\sigma_\mathrm{1D} \gg c_s$ one substitutes $\rho \propto r^{-q}$ into \cref{eq:dimensional_steady_equilibrium}, one can show that the outer envelope of the \ac{TES} becomes a power law with $q \to 2 - 2p$ when $0< p < 0.5$ and $q \to 2p$ when $0.5 \le p < 1$.
The corresponding limits in the polytropic index $\gamma_p \equiv \partial \ln P_\mathrm{eff} / \partial \ln \rho = 1 - 2p/q$ are $\gamma_p \to (1 - 2p) / (1 - p)$ when $0< p < 0.5$ and $\gamma_p \to 0$ when $0.5 \le p < 1$.
The former limit is equivalent to the \ac{SPS} $\rho \propto r^{-q}$ and $P_\mathrm{eff} \propto \rho^{\gamma_p}$ with $q = 2/(2-\gamma_p)$ \citep{mckee99}: the outer envelope of a highly turbulent \ac{TES} with $0 < p < 0.5$ approaches the \ac{SPS} solution.
For $0.5 \le p < 1$, a singular polytropic solution to \cref{eq:dimensional_steady_equilibrium} does not exist.
We note that the convergence of $q \to 1$ when $p=0.5$ is extremely slow: the local slope of the \ac{TES} solution manages to reach $q = 1.05$ at $\xi_\mathrm{crit} = 10^9$.
Because of this, the outer envelope of the $p=0.5$ \ac{TES} shown in \cref{fig:profiles_norm}(c) has not yet reached the power-law limit of $q\to 1$, while the solutions with $p=0.3$ and $0.7$ almost reach the limiting value $q \to 1.4$.

For all these solutions, equilibrium is achieved by balancing the three forces: thermal pressure gradient ($f_\mathrm{thm}$), turbulent pressure gradient ($f_\mathrm{trb}$), and gravity ($f_\mathrm{grv}$) (e.g., \cref{eq:force_balance}).
While $f_\mathrm{thm}$ is always positive and $f_\mathrm{grv}$ negative, the direction of $f_\mathrm{trb}$ can be either outward or inward depending on the sign of $2p - q$, because locally $P_\mathrm{trb} \propto r^{2p - q}$ (see the last paragraph of \cref{sec:equilibrium_sequence}).
\cref{fig:profiles_norm}(d) shows the relative importance of gravity compared to the other two forces in determining the equilibrium, by plotting the quantity
\begin{equation}
    \mathcal{R}_\mathrm{grv} \equiv \frac{|f_\mathrm{grv}|}{\max\left(|f_\mathrm{thm}|, |f_\mathrm{trb}|\right)}
\end{equation}
as a function of radius.
It shows that for all solutions (except the \ac{BE} sphere
where $f_\mathrm{trb} = 0$), $\mathcal{R}_\mathrm{grv} \ll 1$ in the inner part: there, the thermal pressure gradient is balancing the \emph{inward} turbulent pressure gradient, with gravity playing only a minor role.
However, as $q$ increases toward the outer part, the turbulent pressure gradient force decreases in magnitude, and gravity starts to take over.
}

\begin{figure*}[htpb]
  \plotone{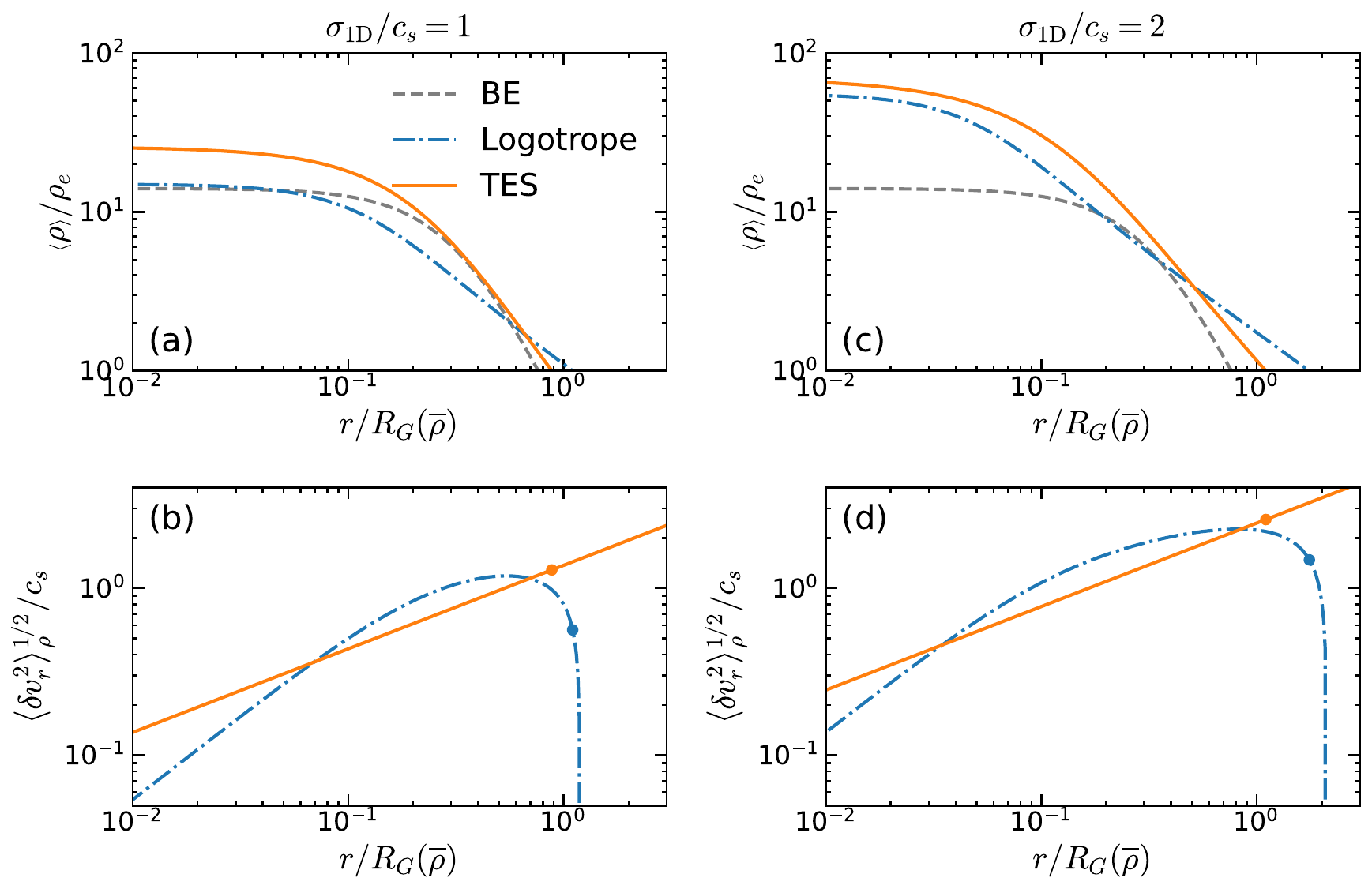}
  \caption{Comparison of the internal structures of the critical \ac{BE} sphere (gray dashed), critical logotrope (blue dot-dashed), and critical \ac{TES} with $p = 0.5$.
    (a) The radial density profile normalized to the edge density. Both the logotrope and \ac{TES} have the mass-weighted average velocity dispersion $\sigma_\mathrm{1D} = c_s$.
    The corresponding parameters are $A = 0.337$ for the logotrope and $\xi_s = 6.42$ for the \ac{TES}.
    (b) The turbulent Mach number versus radius. The locations of critical radii are marked with circles.
    Panels (c) and (d) are similar to (a) and (b), respectively, but with higher velocity dispersion $\sigma_\mathrm{1D} = 2c_s$;
    the corresponding parameters for the logotrope and \ac{TES} are $A = 0.235$ and $\xi_s = 3.42$, respectively.
}
  \label{fig:logotrope}
\end{figure*}

As mentioned in \cref{sec:intro}, a previous theoretical model that was introduced in order to represent nonthermal support was the logotrope \citep{mclaughlin96,mclaughlin97}.   
\cref{fig:logotrope} compares the internal structure of the \ac{TES} with that of the logotrope having the identical velocity dispersion.
It shows that compared to the \ac{TES}, the logotrope has a significantly flatter outer density profile.
\cref{fig:logotrope} also compares the linewidth-size relation for the \ac{TES} with that of the logotrope, where the unphysical turnover of the logotrope is evident at large scales.

\begin{figure*}[htpb]
  \plotone{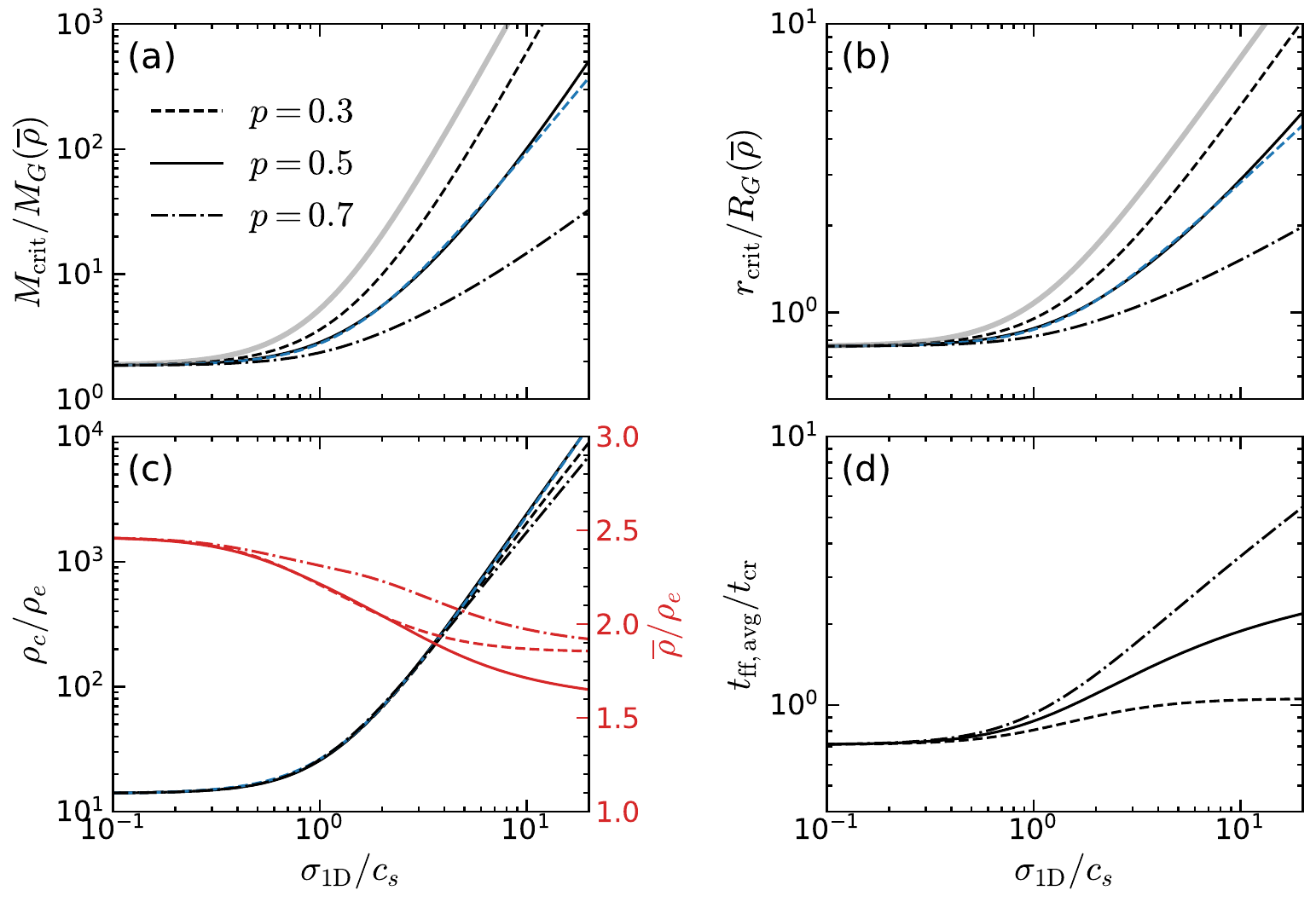}
  \caption{Dependence of physical properties of the critical \ac{TES} on the internal turbulent Mach number $\sigma_\mathrm{1D}/c_s$, for $p=0.3$ (dashed), $0.5$ (solid), and $0.7$ (dot-dashed).
  (a) The critical mass normalized by the gravitational mass $M_G$ (\cref{eq:mgrav}) evaluated at the mean core density $\overline{\rho}=M_\mathrm{crit}/(4\pi r_\mathrm{crit}^3/3)$.
  (b) The critical radius normalized by the gravitational radius $R_G$ (\cref{eq:rgrav}) evaluated at $\bar\rho$.
  (c) The critical center-to-edge density contrast (black, left axis) and the ratio of the mean density to the edge density (red, right axis).
  (d) The ratio between the free-fall time to crossing time at the mean core density.
  \REV{The blue dashed lines in panels (a)--(c) plot analytic approximations given in \cref{eq:mcrit_highsigma,eq:rcrit_highsigma,eq:ucrit_highsigma}.
  The gray bands in panels (a) and (b) plot the \ac{BE} mass and radius resulting from a naive substitution $c_s \to (c_s^2 + \sigma_\mathrm{1D}^2)^{1/2}$ in \cref{eq:mbe,eq:rbe,eq:mgrav,eq:rgrav}.}}
  \label{fig:critical_quantities}
\end{figure*}

Using \cref{eq:def_xi,eq:def_u,eq:enclosed_mass}, the critical mass and radius of a \ac{TES} can be expressed in terms of its center, edge, or mean densities as follows:
\begin{equation}\label{eq:mcrit}
  \begin{split}
    M_\mathrm{crit} &= \frac{m_\mathrm{crit}}{(4\pi)^{1/2}}M_G(\rho_c),\\
                    &= \frac{m_\mathrm{crit}e^{-u_\mathrm{crit}/2}}{(4\pi)^{1/2}}M_G(\rho_e),\\
                    &= \left(\frac{3}{4\pi}\right)^{1/2} \left( \frac{m_\mathrm{crit}}{\xi_\mathrm{crit}} \right)^{3/2}M_G(\overline{\rho}),
  \end{split}
\end{equation}
\begin{equation}\label{eq:rcrit}
  \begin{split}
    r_\mathrm{crit} &= \frac{\xi_\mathrm{crit}}{(4\pi)^{1/2}}R_G(\rho_c),\\
                    &= \frac{\xi_\mathrm{crit}e^{-u_\mathrm{crit}/2}}{(4\pi)^{1/2}}R_G(\rho_e),\\
                    &= \left(\frac{3}{4\pi} \right)^{1/2} \left( \frac{m_\mathrm{crit}}{\xi_\mathrm{crit}} \right)^{1/2}R_G(\overline{\rho}),
  \end{split}
\end{equation}
where $\rho_e \equiv \left<\rho \right>(r = r_\mathrm{crit})$ is the edge density, $\overline{\rho} \equiv 3M_\mathrm{crit}/(4\pi r_\mathrm{crit}^3)=3\rho_c m_\mathrm{crit}/\xi_\mathrm{crit}^3$ is the mean density, and $M_G$ and $R_G$ are the gravitational mass and radius defined in \cref{eq:mgrav,eq:rgrav}.
We remind the reader that $u_\mathrm{crit}$, $\xi_\mathrm{crit}$ and $m_\mathrm{crit}$ are dimensionless functions of $p$ and $\xi_s$, or alternatively, of $p$ and $\sigma_\mathrm{1D}$ (see \cref{fig:critical_tes}(d) for the relationship between $\xi_s$ and $\sigma_\mathrm{1D}$).
We note that \cref{eq:mcrit,eq:rcrit} reduce to \cref{eq:mbe,eq:rbe} in the limit of $\sigma_\mathrm{1D} \to 0$.

\cref{fig:critical_quantities}(a),(b) plots $M_\mathrm{crit}/M_G(\overline{\rho})$ and $r_\mathrm{crit}/R_G(\overline{\rho})$ as functions of $\sigma_\mathrm{1D}/c_s$, showing that both quantities increase with $\sigma_\mathrm{1D}/c_s$, rather slowly when $\sigma_\mathrm{1D} \lesssim c_s$ and rapidly for $\sigma_\mathrm{1D} \gtrsim c_s$.
For $p=0.5$ and $\sigma_\mathrm{1D} = c_s$, we find $r_\mathrm{crit} = 0.88 R_G(\overline{\rho})$ and $M_\mathrm{crit} = 2.86 M_G(\overline{\rho})$, which are factors of $1.15$ and $1.54$ larger than the corresponding values for the \ac{BE} sphere having the same mean density (see also the profile comparison in \cref{fig:profiles_norm}).\footnote{We note that the difference between the \ac{TES} and \ac{BE} sphere is more pronounced when compared at the same central density (see \cref{fig:critical_tes}).}
\REV{\cref{fig:critical_quantities}(a),(b) also plot the ``turbulent \ac{BE}'' mass and radius resulting from a naive substitution $c_s\to (c_s^2 + \sigma_\mathrm{1D}^2)^{1/2}$ in \cref{eq:mbe,eq:rbe,eq:mgrav,eq:rgrav}, which significantly overestimate the true critical mass and radius depending on the Mach number.
For example, the ``turbulent \ac{BE}'' mass and radius
of mildly supersonic cores ($\sigma_\mathrm{1D} = 2c_s$) are larger than $M_\mathrm{crit}$ and $R_\mathrm{crit}$ by a factor of $3.7$ and $1.5$ (assuming $p=0.5$), respectively, indicating such simple substitution would lead to incorrect results.}

When $\sigma_\mathrm{1D} \gg c_s$, the critical radius and mass as well as the critical density contrast all increase as a power-law in $\sigma_\mathrm{1D}/c_s$.
\REV{For example, we find that the critical quantities for $p=0.5$ are well approximated by
\begin{equation}\label{eq:mcrit_highsigma}
  \begin{split}
    M_\mathrm{crit} &\approx M_\mathrm{BE}(\overline{\rho}) \left(1 + \frac{\sigma_\mathrm{1D}^2}{2c_s^2}\right) \\
                    &\approx 30\,M_\odot \left( \frac{T}{20\,\mathrm{K}} \right)^{3/2} \left( \frac{\overline{n}_\mathrm{H}}{500\,\mathrm{cm}^{-3}} \right)^{-1/2}\\
                    &\quad \times \left[ 1 + 64\left(\frac{\sigma_\mathrm{1D}}{3\,\mathrm{km}\,\mathrm{s}^{-1}}\right)^{2}\left(\frac{T}{20\,\mathrm{K}}\right)^{-1} \right],
  \end{split}
\end{equation}
\begin{equation}\label{eq:rcrit_highsigma}
  \begin{split}
    r_\mathrm{crit} &\approx R_\mathrm{BE}(\overline{\rho})\left( 1 + \frac{\sigma_\mathrm{1D}^2}{2c_s^2} \right)^{1/3}\\
                    &\approx 0.74\,\mathrm{pc} \left( \frac{T}{20\,\mathrm{K}} \right)^{1/2} \left( \frac{\overline{n}_\mathrm{H}}{500\,\mathrm{cm}^{-3}} \right)^{-1/2} \\
                    &\quad \times \left[ 1 + 64\left(\frac{\sigma_\mathrm{1D}}{3\,\mathrm{km}\,\mathrm{s}^{-1}}\right)^{2}\left(\frac{T}{20\,\mathrm{K}}\right)^{-1} \right]^{1/3},
  \end{split}
\end{equation}
\begin{equation}\label{eq:ucrit_highsigma}
    \left(\frac{\rho_c}{\rho_e}\right)_\mathrm{crit} \equiv e^{u_\mathrm{crit}} \approx 14\left( 1 + 0.7\frac{\sigma_\mathrm{1D}^2}{c_s^2} \right)^{1.2}
\end{equation}
which are plotted in \cref{fig:critical_quantities}(a)--(c) with blue dashed lines.
We note that \cref{eq:mcrit_highsigma,eq:rcrit_highsigma,eq:ucrit_highsigma} are valid within relative error of $5\%$ for $\sigma_\mathrm{1D} < 9.5 c_s$, $\sigma_\mathrm{1D}<13c_s$, $\sigma_\mathrm{1D}<31c_s$ respectively.
Intriguingly, by evaluating  \cref{eq:mcrit_highsigma} at $\sigma_\mathrm{1D}\gtrsim 3\, \mathrm{km}\,\mathrm{s}^{-1}$, it is readily evident that the critical mass at high Mach number $\sigma_\mathrm{1D} \gg c_s$ becomes comparable to typical masses of star clusters, as will be discussed in \cref{sec:cluster_formation}.}

\cref{fig:critical_quantities}(c) plots the center-to-edge density contrast for marginally stable equilibrium solutions, as a function of $\sigma_\mathrm{1D}/c_s$.
The critical center-to-edge density contrast converges to the well-known value of $14$ in the limit $\sigma_\mathrm{1D} \to 0$, but then monotonically increases with $\sigma_\mathrm{1D}/c_s$ with very weak dependence on $p$, reaching $26$ and $70$ at $\sigma_\mathrm{1D} = c_s$ and $2c_s$, respectively.
We note that for the equilibrium with larger $p$, the outer density profile has to have a steeper slope to generate outward pressure gradient, leading to smaller $r_\mathrm{crit}$ and $M_\mathrm{crit}$ despite having similar center-to-edge density contrast.

\cref{fig:critical_quantities}(c) also shows the ratio of the average density to the edge density (right axis). 
The average density is a factor of $ \sim 1.5 - 2.5$ larger than the edge density, with a mild dependence on $\sigma_\mathrm{1D}/c_s$ and $p$, and the largest ratio for the limiting non-turbulent case.
The ratio $\bar{\rho}/\rho_e$ is only $\sim 1.5-2$ for the highly turbulent cores that have extreme center-to-edge density contrast $\rho_c/\rho_e > 10^2$; while central densities are large, very little mass is involved.

\cref{fig:critical_quantities}(d) plots the ratio of the gravitational free-fall time $t_\mathrm{ff,avg} \equiv \left[3\pi/\left(32 G\overline{\rho}\right)\right]^{1/2}$ at the mean core density to the effective crossing time $t_\mathrm{cr}\equiv r_\mathrm{crit}/(c_s^2 + \sigma_\mathrm{1D}^2)^{1/2}$.
The large $t_\mathrm{ff,avg}/t_\mathrm{cr}$ for $\sigma_\mathrm{1D} \gg c_s$ and small $p$ may indicate that turbulence would quickly rearrange fluid elements in the outer part of the core before gravity can bring about collapse.
For $p<0.5$, however, $t_\mathrm{ff,avg}/t_\mathrm{cr} \sim 0.7 - 2$, so turbulence would be less likely to disrupt cores before the structure as a whole collapses.

\begin{figure}[htpb]
  \plotone{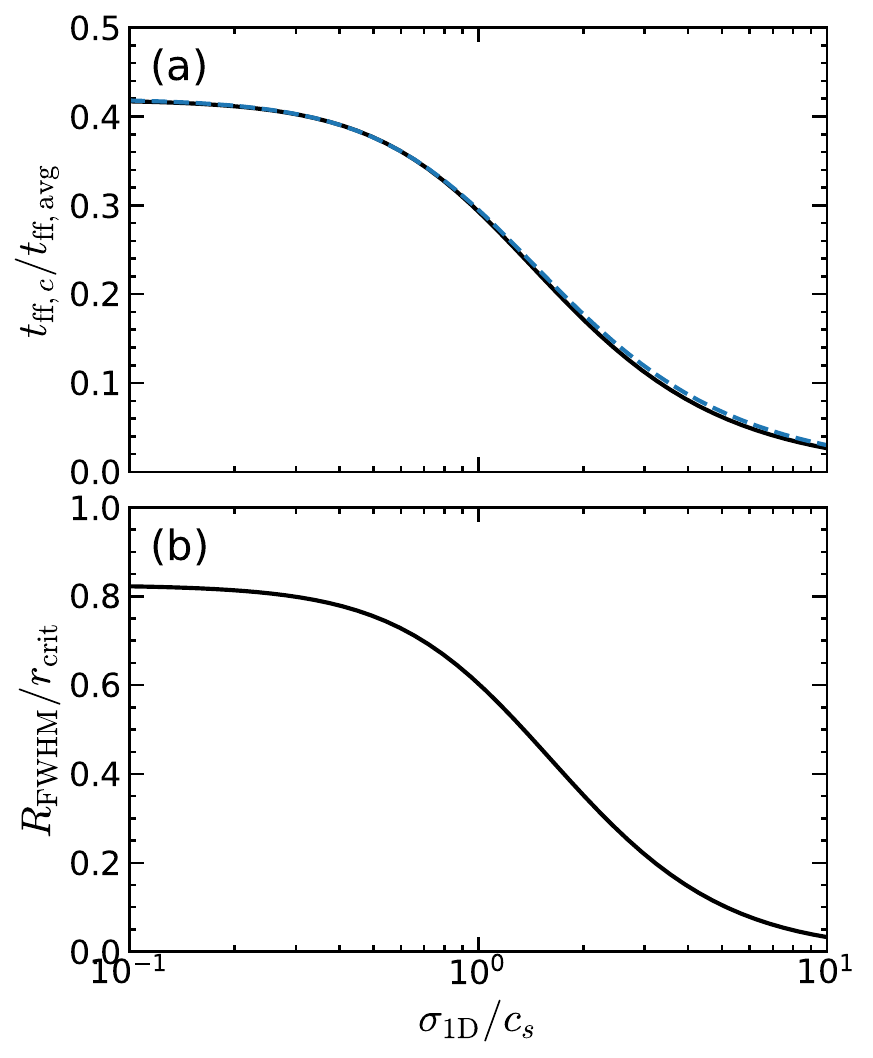}
  \caption{(a) The ratio of the central free-fall time to the free-fall time defined at the average density $\overline{\rho} \equiv 3M_\mathrm{crit}/(4\pi r_\mathrm{crit}^3)$.
  \REV{Blue dashed line plots an analytic approximation provided in \cref{eq:timescale_ratio_analytic}.}
  (b) The ratio of $R_\mathrm{FWHM}$ computed from the column density profile to $r_\mathrm{crit}$.
  Both quantities are plotted against the internal turbulent Mach number $\sigma_\mathrm{1D}/c_s$ of the critical \ac{TES} for turbulent power law index $p=0.5$.}
  \label{fig:critical_tes_obs_properties}
\end{figure}

Observations with limited spatial resolutions are able to constrain $\overline{\rho}$ and therefore $t_\mathrm{ff,avg}$.
However, the characteristic timescale of the evolution leading to the formation of a singularity (i.e., protostar) at the center is $t_{\mathrm{ff},c} \equiv [3\pi / (32 G \rho_c)]^{1/2}$, which is shorter than $t_\mathrm{ff,avg}$.
\REV{Using the data shown in \cref{fig:critical_quantities}(c), we find an analytic approximation
\begin{equation}
    \left(\frac{\rho_c}{\overline{\rho}}\right)_\mathrm{crit} \approx 5.70 \left( 1 + 0.8 \frac{\sigma_\mathrm{1D}^2}{c_s^2}\right)^{1.2}
\end{equation}
analogous to \cref{eq:ucrit_highsigma}, which indicates that the ratio
\begin{equation}\label{eq:timescale_ratio_analytic}
    \frac{t_{\mathrm{ff},c}}{t_\mathrm{ff,avg}} = \left(\frac{\rho_c}{\overline{\rho}}\right)_\mathrm{crit}^{-1/2} \approx 0.42\left(1 + 0.8\frac{\sigma_\mathrm{1D}^2}{c_s^2}\right)^{-0.6}
\end{equation}
is a decreasing function of $\sigma_\mathrm{1D}/c_s$ (see \cref{fig:critical_tes_obs_properties}(a)).
We note that $t_{\mathrm{ff},c}$ is a factor of $\sim 3$ times smaller than $t_\mathrm{ff,avg}$ for a transonically turbulent core.}
Since $t_\mathrm{ff,avg}/t_\mathrm{cr}\sim 1$ for transonic cores, this implies that the center would collapse on a timescale shorter than the outer part would evolve either due to gravity or to turbulence-driven restructuring.
This would be all the more true for highly turbulent cores that reach critical conditions, as $\rho_c/\bar{\rho}\gg 1$ in this limit.

\begin{figure}[htpb]
  \plotone{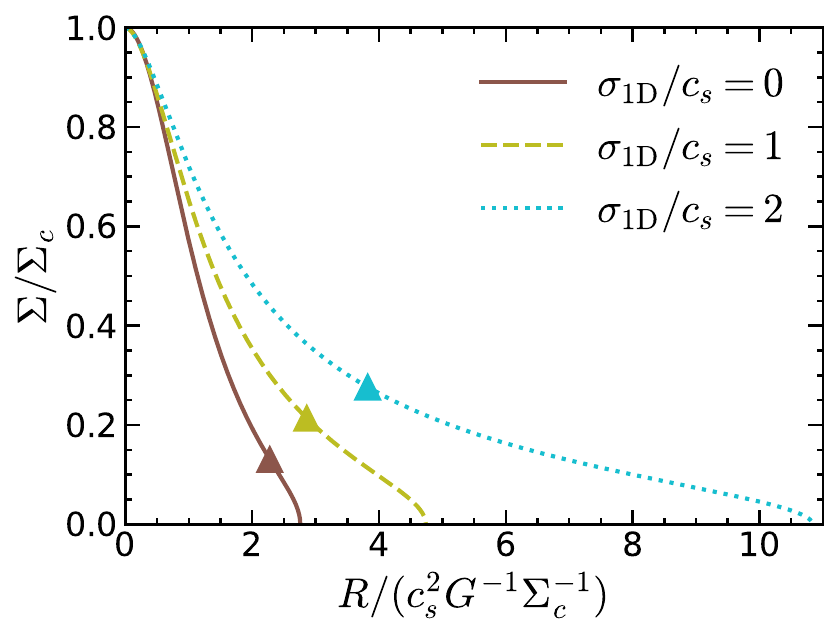}
  \caption{Column density profiles of the critical \acp{TES} with $p=0.5$.
  Solid, dashed, and dotted lines correspond to the internal turbulent Mach number $\sigma_\mathrm{1D}/c_s = 0$, $1$, and $2$, respectively.
  The column density $\Sigma$ is normalized to its central value $\Sigma_c$.
  The profiles extend to $r_\mathrm{crit}$, and the observational core radius $R_\mathrm{FWHM}$ is marked with triangles.}
  \label{fig:column_density_structure}
\end{figure}

Observational sensitivity limits generally make it difficult to probe the outer envelopes of cores.
It is common to define the core radius using the FWHM of the column density profile, which is known to approximate $R_\mathrm{BE}$ if the underlying density distribution is similar to the \ac{BE} sphere \citep[e.g.,][]{konyves15}.
We introduce the notation $R_\mathrm{FWHM}$ to denote this radius, i.e., $R_\mathrm{FWHM} \equiv 2R_\mathrm{HM}$ where the column density at $R_\mathrm{HM}$ is the half of the central value.
To relate $R_\mathrm{FWHM}$ to the critical radius for cores having nonzero velocity dispersion, we first 
integrate to obtain 2D maps of \ac{TES}, thereby obtaining column density profiles.
\cref{fig:column_density_structure} compares examples of the column density profiles for the critical \acp{TES} having zero, transonic, and mildly supersonic turbulent velocity dispersions (see \cref{fig:profiles_norm} for the corresponding volume density profiles normalized to the edge density).
For the latter two, we assume the turbulent power-law index of $p=0.5$.
Based on the column density profiles, we plot the ratio $R_\mathrm{FWHM}/r_\mathrm{crit}$ in \cref{fig:critical_tes_obs_properties}(b).
This shows that while $R_\mathrm{FWHM}/r_\mathrm{crit} \sim 0.8$ when $\sigma_\mathrm{1D} \lesssim 0.5 c_s$, this ratio is smaller by a factor of two or more for cores having supersonic turbulent velocity dispersions.
That is, when the turbulence level is high, the radius of the core must extend well beyond $R_\mathrm{FWHM}$ for a core to be unstable to collapse.
The dotted curve in \cref{fig:column_density_structure} shows an example of this.

\section{Discussion}\label{sec:discussion}

\subsection{Collapse Scenario}\label{sec:collapse_scenario}

In the idealized theory, a quasi-equilibrium core supported by thermal and turbulent pressure is unstable if its maximum radius $r_\mathrm{max}$ exceeds $r_\mathrm{crit}$.
However, the \ac{TES} model describes a single isolated object, without regard to any gravitational forces other than those produced by the core itself.
In reality, hierarchical structure around the core contributes to the landscape of the gravitational potential, and $r_\mathrm{max}$ must also be effectively tidally limited.
For example, in the case of two cores with equal mass, the effective tidal limit is half of their separation.
More generally, if we take $r_\mathrm{max}$ to be the distance to the nearest saddle point of the gravitational potential, it scales linearly with the distance to an external structure of mass $M_\mathrm{external}$, and inversely with $1 + (M_\mathrm{external}/M_\mathrm{core})^{1/2}$.
While making a quantitative prediction of $r_\mathrm{max}$ is beyond the scope of this work, it is clear that the spatial extent of any realistic core forming inside a turbulent cloud is limited by the scale associated with the velocity perturbation creating the core.
In observations, transformation of column density maps of molecular clouds into a dendrogram \citep[e.g.,][]{kirk13} can provide one way of characterizing $r_\mathrm{max}$.

\begin{figure*}[htpb]
  \plotone{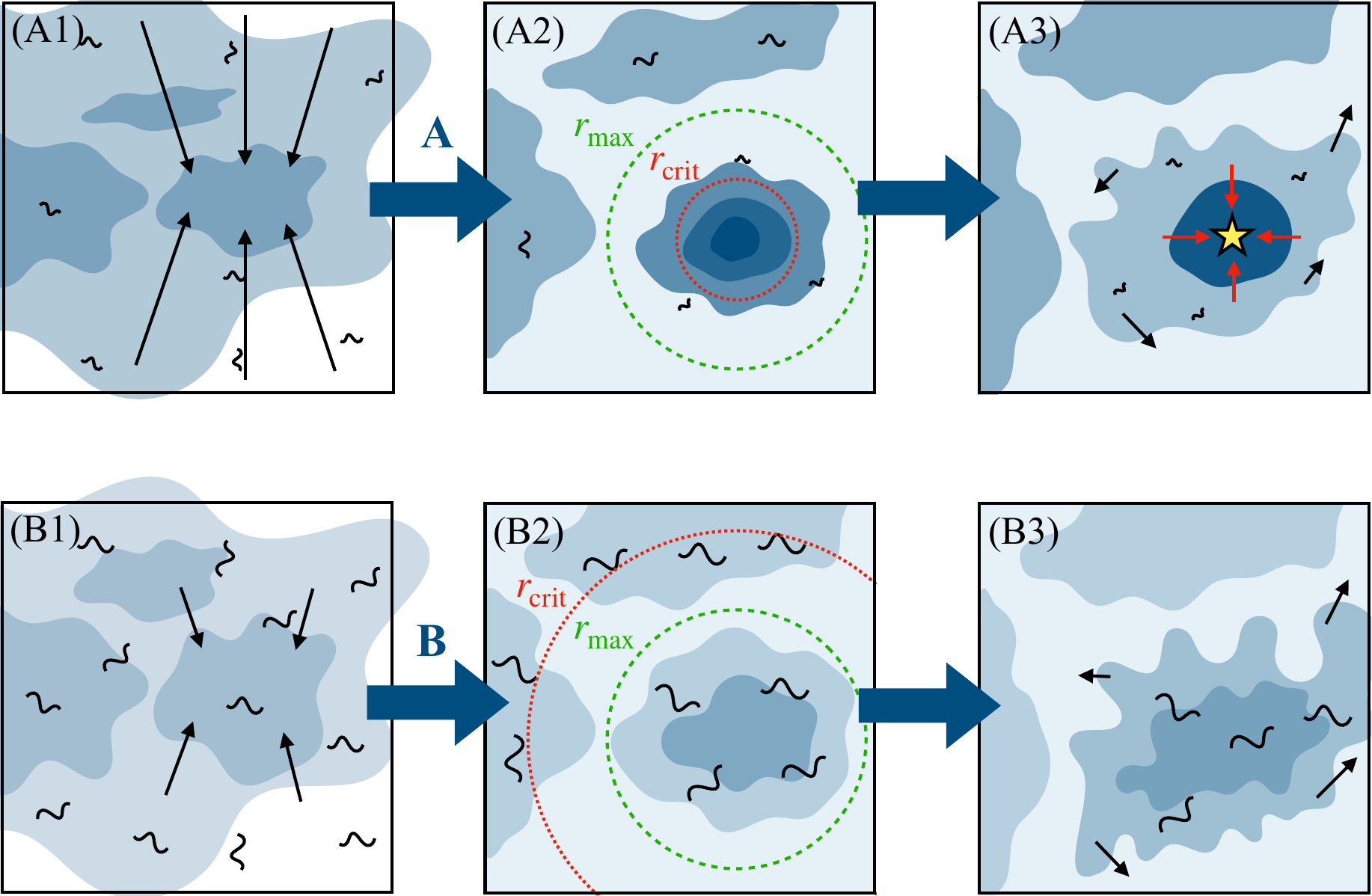}
  \caption{Schematic illustration of two alternative evolutionary scenarios for a nascent core.
  The first scenario leads to collapse/star formation (path \textbf{A}: (A1)$\to$(A2)$\to$(A3)); the second to dispersal (path \textbf{B}: (B1)$\to$(B2)$\to$(B3)).
  (A1) A core begins to form in the region where turbulent flows are locally converging.
  (A2) The forming core becomes dense enough and/or turbulence sufficiently dissipates such that $r_\mathrm{crit}$ becomes smaller than the maximum radius $r_\mathrm{max}$ of a core (beyond which gas belongs to other structures).
  (A3) The core undergoes outside-in collapse as the region near $r_\mathrm{crit}$ becomes unstable.
  (B1) A core begins to form by weakly converging flows.
  (B2) The core approaches quasi-equilibrium, but its critical radius lies outside the maximum radius because of weak initial compression and/or strong local turbulence.
  (B3) The core is subsequently torn apart by turbulence rather than collapsing because the region inside $r_\mathrm{max}$ is stable.}
  \label{fig:evolution_scenario}
\end{figure*}

\cref{fig:evolution_scenario} schematically illustrates two alternative evolutionary scenarios leading to collapse or dispersal.
An overall assumption is that a core is formed by turbulent flows that are locally converging, and the density of the core is determined by the strength of the converging motion.
Larger-amplitude and longer-duration converging velocity perturbations will create higher local density contrast relative to ambient values.
At the same time, turbulence on scales smaller than the overall converging flow will help to protect against collapse by providing support against gravity.

If the converging flow is strong \REV{(creating large $\rho_c)$} and/or the smaller-scale turbulence is weak (i.e., large $r_s$), the resulting core would have large enough $\xi_s$ such that $r_\mathrm{crit} < r_\mathrm{max}$.
The core would then subsequently undergo collapse leading to star formation.
In contrast, when the converging flow is too weak or turbulence is too strong such that $r_\mathrm{crit} > r_\mathrm{max}$, the entire region within $r_\mathrm{max}$ is stable and the core would disperse unless it is further compressed by, e.g., passing shock waves.

In Paper II, we will test the above scenario by directly comparing $r_\mathrm{crit}$ and $r_\mathrm{max}$ for cores forming in a large suite of numerical simulations.
By tracking cores over time, we show that the net radial force within each core becomes negative (i.e., accelerating the inward bulk motions) when $r_\mathrm{crit}$ from \ac{TES} theory first falls below $r_\mathrm{max}$ computed from the full gravitational potential.
We also show that the internal density profiles of the simulated cores at the time they initiate gravitational collapse are consistent with the \ac{TES} solutions obtained in this paper.
We will connect the timescales of different stages of evolution to observed core lifetimes in future work.

\subsection{Projection Effect on Linewidth-Size Relation}\label{sec:projection}

A basic assumption of the \ac{TES} model is that the turbulent velocity dispersion increases with radius as a power law (\cref{eq:linewidth_size}).
Although this assumption is motivated by several observational studies \citep{fuller92,caselli95,choudhury21}, a direct comparison between \cref{eq:linewidth_size} and the observed linewidth-size relation is difficult because the latter involves the integration along the line-of-sight, weighted by the density and emissivity of the transition.
To explore how the intrinsic turbulent velocity structures within a core are related to the observable line-of-sight velocity dispersion, it is necessary to perform numerical simulations in which prestellar cores form within the context of a larger scale \ac{GMC}, which is done in Paper II.
Our preliminary analyses from those simulations show that the line-of-sight non-thermal velocity dispersion traced by dense gas is approximately constant within a core, reminiscent of the so-called ``coherent cores'' \citep[e.g.,][]{goodman98,chen19}, even though the intrinsic velocity dispersion obeys a power-law.
Careful modeling including density selection effect of different molecular tracers would be required to relate the observed linewidth-size relation to the intrinsic turbulent structures in three dimensions.

\subsection{Implication for Star Cluster Formation}\label{sec:cluster_formation}

Massive star clusters such as those found within Carina or the Orion Nebula contain a few thousand solar masses within a few parsecs and have velocity dispersions of a few kilometers per second \citep{dario14,shull21,theissen22}.
Given that the thermal Jeans mass under  \ac{GMC} conditions on large scales is $\sim 10-10^2\,M_\odot$, a longstanding question is how material can avoid early fragmentation prior to assembling the massive, compact structures that are cluster progenitors.
Our \ac{TES} solutions shed some light on this question: provided that the central density remains lower than the value given in \cref{eq:rho_min} or \cref{eq:nc_min}, a turbulent overdense patch of a GMC \REV{encompassing significant mass} could in principle become progressively more concentrated without undergoing runaway collapse. 
Assuming $T = 20\,\mathrm{K}$, $\overline{n}_\mathrm{H} = 500\,\mathrm{cm}^{-3}$, and $\sigma_\mathrm{1D} = 3\,\mathrm{km}\,\mathrm{s}^{-1}$, typical of dense clumps in GMCs, \cref{eq:mcrit_highsigma,eq:rcrit_highsigma} yield $M_\mathrm{crit}  \sim 2\times 10^3\,M_\odot$ and $r_\mathrm{crit} \sim 3\,\mathrm{pc}$, which are intriguingly similar to typical mass and radius of ONC-like clusters.
The quantitative results from our \ac{TES} model may therefore be key to realizing the equilibrium cluster formation scenario of \citet{tan06}.

\begin{acknowledgments}
This work was supported in part by grant 510940 from the Simons Foundation to E.~C.\ Ostriker.
Computational resources for this project were provided by Princeton Research Computing, a consortium including PICSciE and OIT at Princeton University.
\REV{We are grateful to the anonymous reviewer for constructive comments on the manuscript.}
\end{acknowledgments}

\bibliographystyle{aasjournal}
\bibliography{mybib}

\appendix

\crefalias{section}{appsec}

\section{Alternative Formulation}\label{app:alt_formulation}

In this appendix, we provide an alternative nondimensionalization of \cref{eq:dimensional_steady_equilibrium} based on the external pressure $P_\mathrm{ext}$ rather than the central density.
We find this formulation useful when probing the highly turbulent regime where the dependence of the critical quantities on $\xi_s$ becomes extremely sensitive (see \cref{fig:critical_tes}).
In order to distinguish the alternative dimensionless variables from those in the main text, we attach a prime to each variable and define
\begin{align}
  \xi' &\equiv \frac{G^{1/2}P_\mathrm{ext}^{1/2}}{c_s^2}r,\\
  u'(\xi';\xi'_s,p) &\equiv \ln \frac{P_\mathrm{eff}}{P_\mathrm{ext}},\\
  m'(\xi';\xi'_s,p) &\equiv \frac{G^{3/2}P_\mathrm{ext}^{1/2}}{c_s^4} M_\mathrm{enc},\\
  \chi'(\xi';\xi'_s,p) &\equiv 1 + \left( \frac{\xi'}{\xi'_s} \right)^{2p}. 
\end{align}
In terms of these dimensionless variables, \cref{eq:dimensional_steady_equilibrium} becomes
\begin{equation}\label{eq:emden_alt}
  \frac{1}{\xi'^2} \frac{\partial}{\partial\xi'} \left( \chi' \xi'^2 \frac{\partial u'}{\partial \xi'} \right) = - \frac{4\pi e^{u'}}{\chi'}.
\end{equation}
Starting from very small $\xi'$, one can integrate \cref{eq:emden_alt} in terms of the logarithmic radius $t' \equiv \ln \xi'$ with the initial conditions $u' = u_c$ and $\partial u' / \partial t = 0$, until $P_\mathrm{eff}$ matches the external pressure $P_\mathrm{ext}$ (i.e., $u' = 0$) at the outer radius $\xi_\mathrm{ext}$.
With all other parameters fixed, the total mass of the sphere $m'(\xi_\mathrm{ext})$ initially increases with $u_c$ (starting from $u_c = 0$), and then reaches the maximum when $u_c = u_{c,\mathrm{crit}}$.
It is not difficult to see that this stationary point is equivalent to the point where $K = 0$ and thus defines the marginally stable solution.
This also implies that, for a given external pressure, there is a maximum mass above which no equilibrium solution exists.

\end{document}